\documentclass[twocolumn,showpacs,preprintnumbers,amsmath,amssymb,prd, 
superscriptaddress]{revtex4-1}

\usepackage[dvipdfm]{graphicx}
\usepackage{dcolumn}
\usepackage{bm}
\usepackage{lineno}
\usepackage{amsmath}
\usepackage{amssymb}
\usepackage{relsize}
\usepackage{tabularx} 
\begin{document}

\title{Probing the origin of cosmic-rays with extremely high energy neutrinos 
using the IceCube Observatory}

\affiliation{III. Physikalisches Institut, RWTH Aachen University, D-52056 
Aachen, Germany}
\affiliation{School of Chemistry \& Physics, University of Adelaide, 
Adelaide SA, 5005 Australia}
\affiliation{Dept.~of Physics and Astronomy, University of Alaska Anchorage, 
3211 Providence Dr., Anchorage, AK 99508, USA}
\affiliation{CTSPS, Clark-Atlanta University, Atlanta, GA 30314, USA}
\affiliation{School of Physics and Center for Relativistic Astrophysics, 
Georgia Institute of Technology, Atlanta, GA 30332, USA}
\affiliation{Dept.~of Physics, Southern University, Baton Rouge, LA 70813, USA}
\affiliation{Dept.~of Physics, University of California, Berkeley, CA 94720, 
USA}
\affiliation{Lawrence Berkeley National Laboratory, Berkeley, CA 94720, USA}
\affiliation{Institut f\"ur Physik, Humboldt-Universit\"at zu Berlin, D-12489 
Berlin, Germany}
\affiliation{Fakult\"at f\"ur Physik \& Astronomie, Ruhr-Universit\"at Bochum, 
D-44780 Bochum, Germany}
\affiliation{Physikalisches Institut, Universit\"at Bonn, Nussallee 12,
 D-53115 Bonn, Germany}
\affiliation{Universit\'e Libre de Bruxelles, Science Faculty CP230, B-1050 
Brussels, Belgium}
\affiliation{Vrije Universiteit Brussel, Dienst ELEM, B-1050 Brussels, Belgium}
\affiliation{Dept.~of Physics, Chiba University, Chiba 263-8522, Japan}
\affiliation{Dept.~of Physics and Astronomy, University of Canterbury, 
Private Bag 4800, Christchurch, New Zealand}
\affiliation{Dept.~of Physics, University of Maryland, College Park, MD 20742, 
USA}
\affiliation{Dept.~of Physics and Center for Cosmology and Astro-Particle 
Physics, Ohio State University, Columbus, OH 43210, USA}
\affiliation{Dept.~of Astronomy, Ohio State University, Columbus, OH 43210, 
USA}
\affiliation{Niels Bohr Institute, University of Copenhagen, DK-2100 
Copenhagen, Denmark}
\affiliation{Dept.~of Physics, TU Dortmund University, D-44221 Dortmund, 
Germany}
\affiliation{Dept.~of Physics, University of Alberta, Edmonton, Alberta, 
Canada T6G 2E1}
\affiliation{Erlangen Centre for Astroparticle Physics, 
Friedrich-Alexander-Universit\"at Erlangen-N\"urnberg, D-91058 Erlangen, 
Germany}
\affiliation{D\'epartement de physique nucl\'eaire et corpusculaire, 
Universit\'e de Gen\`eve, CH-1211 Gen\`eve, Switzerland}
\affiliation{Dept.~of Physics and Astronomy, University of Gent, B-9000 Gent, 
Belgium}
\affiliation{Dept.~of Physics and Astronomy, University of California, Irvine, 
CA 92697, USA}
\affiliation{Laboratory for High Energy Physics, \'Ecole Polytechnique 
F\'ed\'erale, CH-1015 Lausanne, Switzerland}
\affiliation{Dept.~of Physics and Astronomy, University of Kansas, Lawrence, 
KS 66045, USA}
\affiliation{Dept.~of Astronomy, University of Wisconsin, Madison, WI 53706, 
USA}
\affiliation{Dept.~of Physics and Wisconsin IceCube Particle Astrophysics 
Center, University of Wisconsin, Madison, WI 53706, USA}
\affiliation{Institute of Physics, University of Mainz, Staudinger Weg 7, 
D-55099 Mainz, Germany}
\affiliation{Universit\'e de Mons, 7000 Mons, Belgium}
\affiliation{T.U. Munich, D-85748 Garching, Germany}
\affiliation{Bartol Research Institute and Department of Physics and Astronomy,
 University of Delaware, Newark, DE 19716, USA}
\affiliation{Dept.~of Physics, University of Oxford, 1 Keble Road, Oxford OX1 
3NP, UK}
\affiliation{Dept.~of Physics, University of Wisconsin, River Falls, WI 54022, 
USA}
\affiliation{Oskar Klein Centre and Dept.~of Physics, Stockholm University, 
SE-10691 Stockholm, Sweden}
\affiliation{Department of Physics and Astronomy, Stony Brook University, 
Stony Brook, NY 11794-3800, USA}
\affiliation{Department of Physics, Sungkyunkwan University, Suwon 440-746, 
Korea}
\affiliation{Department of Physics, University of Toronto, Toronto, Ontario, 
Canada, M5S 1A7}
\affiliation{Dept.~of Physics and Astronomy, University of Alabama, Tuscaloosa,
AL 35487, USA}
\affiliation{Dept.~of Astronomy and Astrophysics, Pennsylvania State 
University, University Park, PA 16802, USA}
\affiliation{Dept.~of Physics, Pennsylvania State University, University Park, 
PA 16802, USA}
\affiliation{Dept.~of Physics and Astronomy, Uppsala University, Box 516, 
S-75120 Uppsala, Sweden}
\affiliation{Dept.~of Physics, University of Wuppertal, D-42119 Wuppertal, 
Germany}
\affiliation{DESY, D-15735 Zeuthen, Germany}

\author{M.~G.~Aartsen}
\affiliation{School of Chemistry \& Physics, University of Adelaide, Adelaide 
SA, 5005 Australia}
\author{R.~Abbasi}
\affiliation{Dept.~of Physics and Wisconsin IceCube Particle Astrophysics 
Center, University of Wisconsin, Madison, WI 53706, USA}
\author{M.~Ackermann}
\affiliation{DESY, D-15735 Zeuthen, Germany}
\author{J.~Adams}
\affiliation{Dept.~of Physics and Astronomy, University of Canterbury, 
Private Bag 4800, Christchurch, New Zealand}
\author{J.~A.~Aguilar}
\affiliation{D\'epartement de physique nucl\'eaire et corpusculaire, 
Universit\'e de Gen\`eve, CH-1211 Gen\`eve, Switzerland}
\author{M.~Ahlers}
\affiliation{Dept.~of Physics and Wisconsin IceCube Particle Astrophysics 
Center, University of Wisconsin, Madison, WI 53706, USA}
\author{D.~Altmann}
\affiliation{Erlangen Centre for Astroparticle Physics, 
Friedrich-Alexander-Universit\"at Erlangen-N\"urnberg, D-91058 Erlangen, 
Germany}
\author{C.~Arguelles}
\affiliation{Dept.~of Physics and Wisconsin IceCube Particle Astrophysics 
Center, University of Wisconsin, Madison, WI 53706, USA}
\author{J.~Auffenberg}
\affiliation{Dept.~of Physics and Wisconsin IceCube Particle Astrophysics 
Center, University of Wisconsin, Madison, WI 53706, USA}
\author{X.~Bai}
\thanks{Physics Department, South Dakota School of Mines and Technology, 
Rapid City, SD 57701, USA}
\affiliation{Bartol Research Institute and Department of Physics and Astronomy,
University of Delaware, Newark, DE 19716, USA}
% ...

\date{\today}

\collaboration{IceCube Collaboration}
\noaffiliation

\begin{abstract}
\vspace{5mm} 
We have searched for extremely high energy neutrinos using data taken with 
the IceCube detector between May 2010 and May 2012. Two neutrino induced 
particle shower events with energies around 1~PeV were observed, as reported 
previously. In this work, we investigate whether these events could originate 
from cosmogenic neutrinos produced in the interactions of ultra-high energy 
cosmic-rays with ambient photons while propagating through intergalactic space.
Exploiting IceCube's large exposure for extremely high energy neutrinos and 
the lack of observed events above 100~PeV, we can rule out the corresponding 
models at more than 90\% confidence level. 
The model independent quasi-differential 90\% CL upper limit, which 
amounts to $E^2 \phi_{\nu_e + \nu_\mu + \nu_\tau} = 1.2 \times 10^{-7}$ 
GeV cm$^{-2}$ s$^{-1}$ sr$^{-1}$ at 1~EeV,
provides the most stringent constraint in the energy range from 10~PeV to 
10~EeV. Our observation disfavors strong cosmological evolution of the 
highest energy cosmic ray sources such as the Fanaroff-Riley type II class 
of radio galaxies. 
\end{abstract}


\maketitle

\section{Introduction}
\label{Introduction}%
Cosmic neutrinos are expected to be produced in the interactions of high 
energy hadronic particles from cosmic accelerators with surrounding photons 
and matter. 
At PeV energies or greater, neutrinos are a unique tool for the direct 
survey of the ultra-high energy universe, because photons at these energies 
are highly attenuated by the cosmic microwave background (CMB). 
In addition to neutrinos directly produced in cosmic-ray sources, secondary 
neutrinos produced in the propagation of ultra-high energy cosmic-rays 
(UHECRs) with energies reaching about $100\,{\rm EeV}$ are expected. 
These ``cosmogenic'' neutrinos are produced by the Greisen-Zatsepin-Kuzmin 
(GZK) mechanism via
interactions of UHECRs with the CMB and extragalactic background light 
(infra-red, optical, 
and ultra violet)~~\cite{greisen,zatsepin,Beresinsky:1969qj}. 
A measurement of cosmogenic (or GZK) neutrinos probes the origin of the UHECRs
because the spectral shapes and flux levels are sensitive to the redshift 
dependence of UHECR source distributions and cosmic-ray primary 
compositions~\cite{yoshida2012,ahlers2009}. 
Neutrinos are ideal particles to investigate the origin of UHECRs since 
neutrinos propagate to the Earth 
essentially without deflection and absorption.
The main energy range of the cosmogenic neutrinos is predicted to be around 
100~PeV--10~EeV~\cite{yoshida93,ESS}. 
In this extremely high energy (EHE) region, cosmogenic production is 
considered the main source of cosmic neutrinos.

A measurement of these EHE neutrinos requires a detection volume on the 
order of at least 1 km$^3$
as their fluxes are expected to be very low, yielding approximately one 
event per year in such a volume~\cite{juliet, icecubeEHE2011}.
The IceCube Neutrino Observatory~\cite{icecube} at the geographical South 
Pole is the first cubic-kilometer scale neutrino detector. Its large 
instrumented volume as well as its omni-directional neutrino detection 
capability have increased the sensitivity for EHE cosmogenic neutrinos 
significantly.
Previous EHE neutrino searches performed with 
IceCube~\cite{icecubeEHE2010, icecubeEHE2011}, showed that IceCube has 
become the most sensitive neutrino detector in the energy range of 
1~PeV--10~EeV compared to experiments using other 
techniques~\cite{auger09, auger11, anita2, anita2err, rice06}.
The sensitivity of the complete IceCube detector reaches to the modestly 
high flux cosmogenic models which assume a pure proton composition of 
cosmic rays.
The flux for a heavier composition such as iron is at least
2--3 times lower, although the decrease depends on the source 
evolution~\cite{kotera2010} 
and strongly on the maximal injection energy of the 
sources~\cite{Ahlers:2012rz}.
In order to test the heavier composition model predictions, longer exposure or
other detection techniques such as the radio detection are needed.

The EHE neutrino search presented here
uses data obtained from May 2010 to May 2012. The analysis is sensitive to all
three neutrino flavors. The basic search strategies are similar to previous 
searches~\cite{icecubeEHE2010, icecubeEHE2011}. The main improvement comes 
from the enlargement of the detector and the statistical enhancement of the 
data as well as improved modeling of optical properties of the deep glacial 
ice~\cite{Aartsen:2013rt} in the Monte Carlo simulations. The improvements 
allow a refined geometrical reconstruction of background events and thus a 
better background rejection. Two neutrino-induced PeV-energy particle shower 
events were discovered by 
this EHE neutrino analysis as reported in Ref.~\cite{Aartsen:2013bka}. 
In this paper, we describe the details of the analysis.
Then, we investigated
whether the two observed events are consistent with cosmogenic neutrinos.
Afterwards, cosmogenic neutrino models were tested for compatibility with 
our observation in order to constrain the UHECR origin. 

The paper is structred as follows:
in sections \ref{sec:detector} and \ref{sec:datasimulation}, the IceCube 
detector and the data samples are described. 
The improved analysis methods and the associated systematic uncertainties are 
discussed in sections \ref{sec:eventselections} and \ref{sec:Sys}. 
In section~\ref{sec:Resuts}, results from the analysis are presented. 
Implications of the observational results on the UHECR origin are discussed 
in section~\ref{sec:ModelTests} by testing several cosmogenic neutrino models.
The model independent upper limit of the EHE neutrino flux is shown in 
section~\ref{sec:ModelIndepUpLimit}. 
Finally, the results are summarized in section~\ref{sec:summary}.

\section{The IceCube Detector}\label{sec:detector}

\begin{figure}[t]
  \hspace{-2.1cm}
  \includegraphics[width=4.2in]{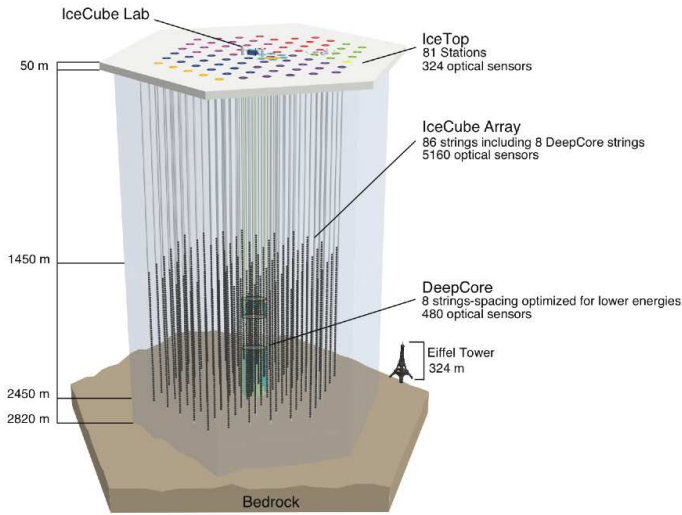}
  \caption{(color online). A schematic view of the IceCube detector.}
\label{fig:schetch}
\end{figure}

The IceCube detector observes the Cherenkov light from the relativistic charged
particles produced by high-energy neutrino interactions, using an array of 
Digital Optical Modules (DOMs).
Each DOM comprises a 10" R7081-02 photomultiplier tube (PMT)~\cite{pmt} in a 
transparent pressure sphere along with a high voltage system, a digital 
readout board~\cite{digitizer}, and
an LED flasher board for optical calibration in ice.
These DOMs are deployed along electrical cable bundles that carry power and 
information 
between the DOMs and the surface electronics. The cable assemblies, called 
strings, were lowered into holes drilled to a depth of 2450~m with an 
horizontal spacing of approximately 125~m (Fig.~\ref{fig:schetch}).
The DOMs sit where the glacial ice is transparent at depths from 1450\,m to 
2450\,m at intervals of 17\,m.
PMT waveforms are recorded when the signal in a DOM crosses a threshold and 
the nearest or next-to-nearest DOM observes a photon within 1\,$\mu$s 
(hard local coincidence, HLC). An event is  triggered if eight DOMs recorded 
an HLC within 5\,$\mu$s.
The lower, inner part of the detector, called DeepCore~\cite{deepcore}, is 
filled with DOMs with a smaller vertical and horizontal spacing of 7 m and 
72 m, respectively. 
The DeepCore array is mainly responsible for the enhancement of the 
performance below 100 GeV, the threshold energy of IceCube. 
Additional DOMs frozen into tanks located at the surface near the top of 
each hole constitute an air shower array called IceTop~\cite{icetop}. 
IceTop allows studying cosmic ray physics and provides the capability to 
study the atmospheric muon background. The whole detector system comprises 
5160 DOMs on 86 strings out of which 8~strings correspond to DeepCore, 
and an additional 324~DOMs in the surface array. The configurations of the 
IceCube detectors are displayed in Fig.~\ref{fig:schetch}.

\section{Data and Simulation}\label{sec:datasimulation}
The IceCube detector construction was completed in December, 2010. 
During the construction phase, from May 31, 2010 to May 12, 2011, 79 strings 
(IC79), approximately 90\% of the full detector, were operational. 
The IC79 run was immediately followed by the first year of data taking with 
the full detector (IC86) which lasted from May 13, 2011 to May 15, 2012. 
The data from these periods were used in this analysis.
The corresponding livetime for the IC79 and IC86 runs are 319.2~days and 
350.9~days respectively, excluding periods of detector calibration and 
unstable operation.
Approximately 10\% of the sample (33.4~days of IC79 and 20.8 days of 
IC86 running) 
was used as a statistically independent test sample for verification. 
The final analysis was performed in a blind way where the test sample was not 
used for the signal search.

There are two classes of background events; atmospheric muon bundle events 
and events induced by atmospheric neutrinos.
Muon bundles consist of a large number of high energy muons produced by 
cosmic-ray interactions in the atmosphere.
Regardless of their high muon multiplicities,  they are observed as a single 
track 
since their lateral separations of about 10 m is shorter than the minimum DOM 
separation of 17 m except for DeepCore.
Since the detector is large and the data recording time window is also long 
(10 $\mu s$), there is a 
non-negligible chance that two or more muon bundles arrive at same time.
These events, called ``coincident events'', complicate geometrical 
reconstruction.
Special treatment is required to reduce this background.
Atmospheric muon bundles were simulated with the CORSIKA air-shower 
simulation~\cite{Heck:1998vt} with the SIBYLL 2.1 hadronic interaction 
model~\cite{Ahn:2009wx}.
Muons from the showers were propagated from the Earth's surface to IceCube 
depths with the MMC package~\cite{mmc}. 
These are the same programs as in previous studies \cite{icecubeEHE2011} 
except that we have improved our description of the optical properties of 
the glacier ice~\cite{Aartsen:2013rt} used in the simulation of the 
photon propagation from the particles to the DOMs. 

For the atmospheric neutrinos, the All Neutrino Interaction Simulation 
(ANIS) package~\cite{anis} is used to simulate each neutrino flavor 
separately between 50~GeV and 1~EeV. 
The neutrino events were simulated following an $E_\nu^{-1}$ 
spectrum on the surface 
of the Earth with appropriate flux weights to represent the spectrum 
resulting from decays of cosmic-ray induced pions and kaons in the atmosphere
(``conventional'' atmospheric neutrinos). We use the cosmic ray spectrum 
modeled in Ref.~\cite{TG}
to take into account the spectral bend at the cosmic ray knee.
The neutrino multiplicity employed in this calculation
is derived from a modified Elbert formula~\cite{TGbook1990,elbert79}.
At PeV energies and above, ``prompt'' atmospheric neutrinos from decays of 
charmed mesons are expected to 
dominate over the conventional atmospheric neutrinos. 
We consider the default value of the prompt neutrino flux from Enberg 
{\it et al.} \cite{prompt_enberg} modified 
to incorporate the cosmic-ray spectrum model in Ref. \cite{TG}.

In order to efficiently simulate high energy events with energies exceeding 
100~TeV at IceCube depths,
the JULIeT package is used in which the propagation of neutrinos is 
efficiently obtained by solving numerical transport equations as described 
in Ref.~\cite{juliet}.

\begin{figure}[t]
  \includegraphics[width=1.5in]{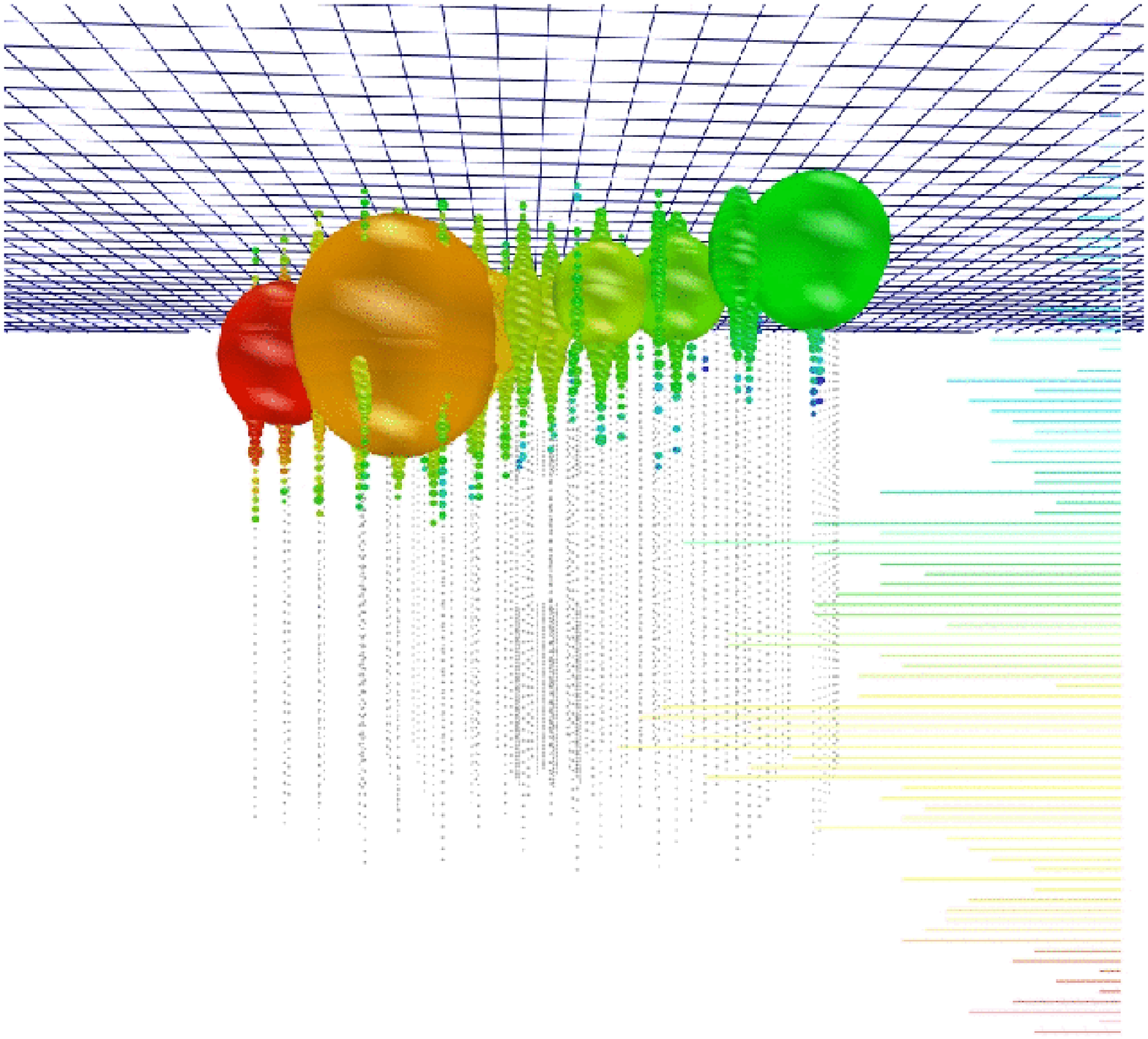}
  \includegraphics[width=1.53in]{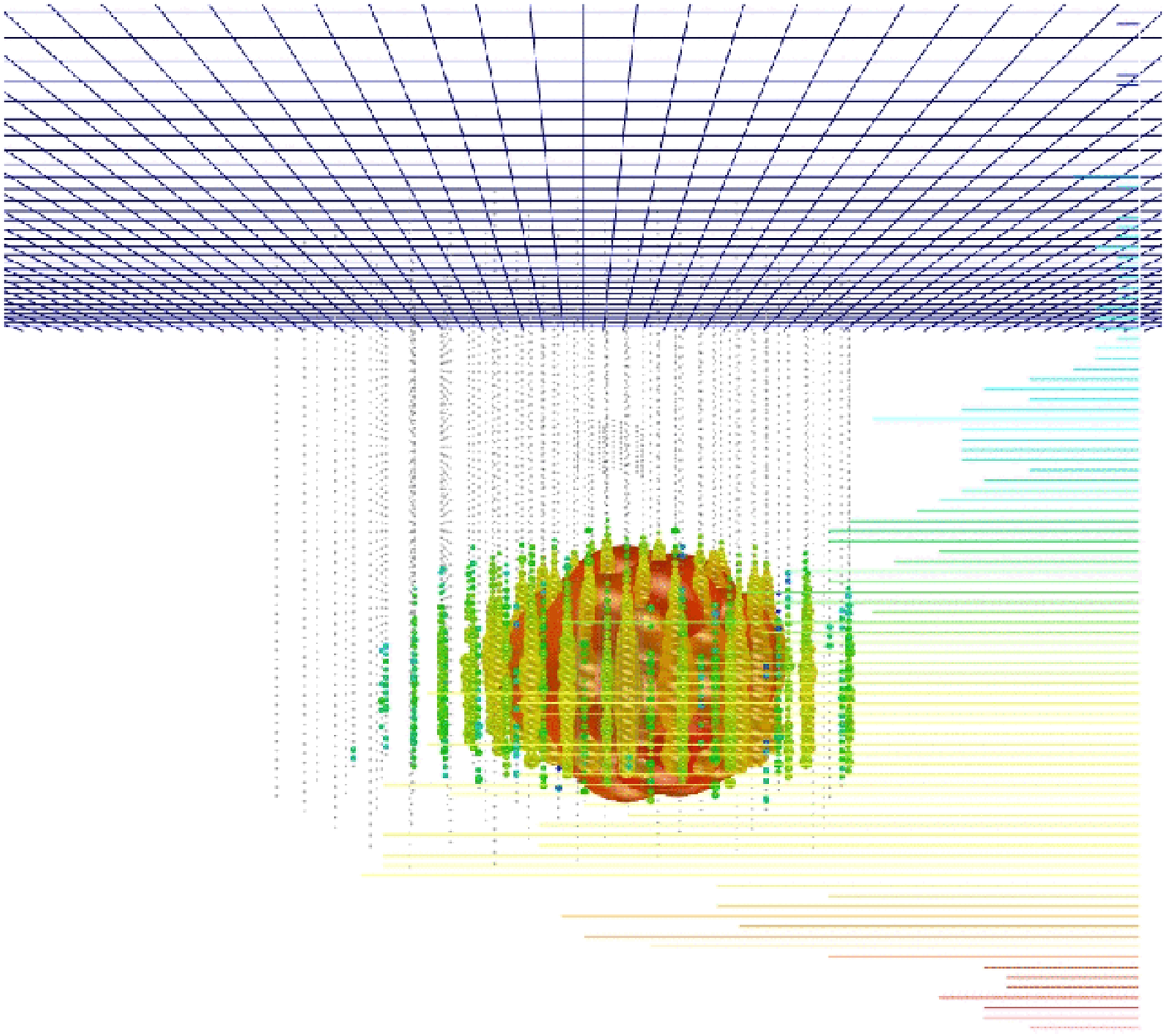}\\  
  \includegraphics[width=1.5in]{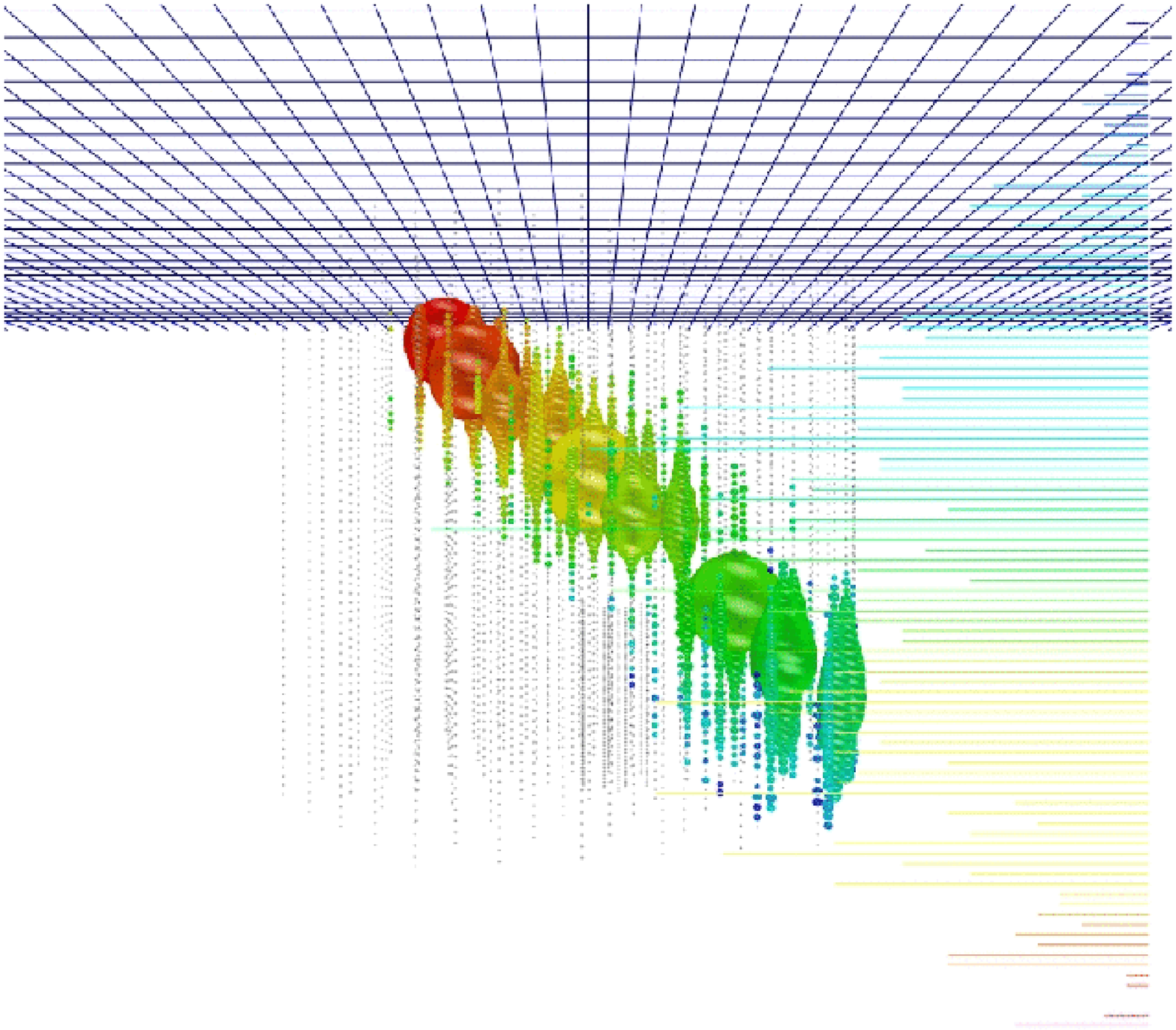}
  \caption{(color online). Event displays of simulated events. 
    Each sphere represents a DOM. 
    Colors indicate the arrival time of the photon 
    (red indicates the earliest and blue the latest). 
    The size of the sphere and the length of the horizontal lines at the 
    right border
    indicate the measured amount of photo-electrons in each DOM.
    Upper left: An upgoing muon entering into the detector array with energy 
    of 20~PeV induced by a neutrino of 500~PeV.
    Upper right: A 300~PeV $\nu_e$ induced cascade event.
    Lower panel: A typical background atmospheric 
    muon bundle event in the current analysis induced by primary cosmic-ray 
    energy of 1~EeV.
    \label{fig:eventviews}
}\end{figure}

Fig.~\ref{fig:eventviews} shows examples of simulated signal and background 
events observed in the IceCube detector. The sizes and colors of the spheres 
indicate the number and the timing of photo-electrons observed in each DOM. 
A signal muon event produces a number of stochastic energy losses along the 
path.
Tau events with energies greater than 10~PeV resemble muon tracks, 
except that they exhibit less energy loss due to their heavier masses. 
They may also generate characteristic ``double bang events'' at energies 
between 1 and 10~PeV due to 
neutrino interactions and successive tau decays inside the detector volume. 
Particle showers are induced by neutral current interactions of neutrinos of 
any flavor or by charged current interactions of electron neutrinos. 
These events, called cascade events, generate spherical hit patterns in the 
detector. A background muon bundle event in the current study typically 
contains 
about 100 to 1000 muons with lateral separations of about 10 m which results 
in a smoother energy loss profile compared to one from a single muon or tau 
event.

\section{Event selections}\label{sec:eventselections}

The energy spectrum of atmospheric muons and neutrinos falls steeply with 
energy. 
The cosmogenic neutrino fluxes with their harder spectra are expected to 
dominate over this background at high energies. 
Because the amount of deposited energy, i.e.~the observable energy, is 
correlated with the energy of the incoming particles, the signal events 
stand out against the background events at high energy.
Therefore, this analysis is targeted towards the selection of these high 
energy events.

The initial event filter selects events containing 
more than 1,000 photo-electrons (p.e.). 
This filtering eliminates a large number of low-energy atmospheric muon 
induced events, typically with less than a few TeV energy.
The filtering process is performed at the South Pole and the resulting 
EHE sample is sent to the North via satellite. 
The samples contained a total of 4.0$\times 10^7$ and 6.0$\times 10^7$ 
events for IC79 and IC86, respectively.

The EHE sample transferred to the North is subjected to off-line hit cleaning
in order to remove coincident atmospheric muons and PMT noise.
A hit represents a reconstructed pulse of photons from a waveform recorded 
by a DOM, 
and is characterized by its time and charge.
The initial hit cleaning is a time window cut on the hits outside the time 
interval between $-4.4\,\mu$s and $+$6.4\,$\mu$s relative to the time of the 
first hit on the DOM with the highest charge. 
Then a secondary hit cleaning based on distances and hit time intervals 
between DOMs is applied. 
Hits from the DeepCore strings are discarded at this stage and not used for 
higher selection levels to keep the DOM separation uniform across the 
detector volume. After these hit cleanings, the analysis level sample was 
selected by requesting at least 300 hits and 3200~p.e. in the whole detector 
except DeepCore.
This sample contains a total of $4.5\times 10^5$ and $5.9\times 10^5$ 
observed events for IC79 and IC86, respectively. 
\begin{figure}[tbp]
  \includegraphics[width=1.68in]{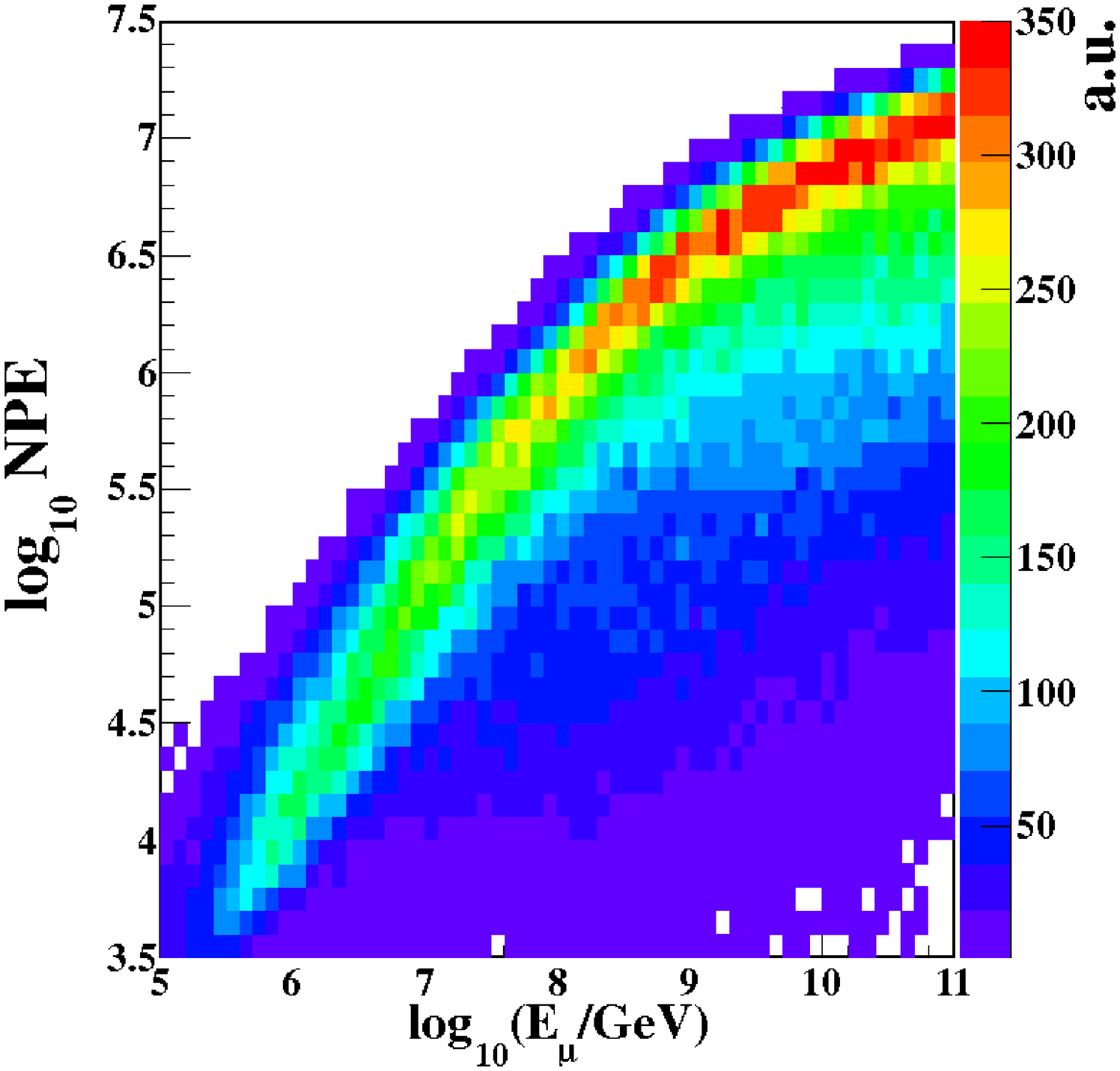}
  \includegraphics[width=1.68in]{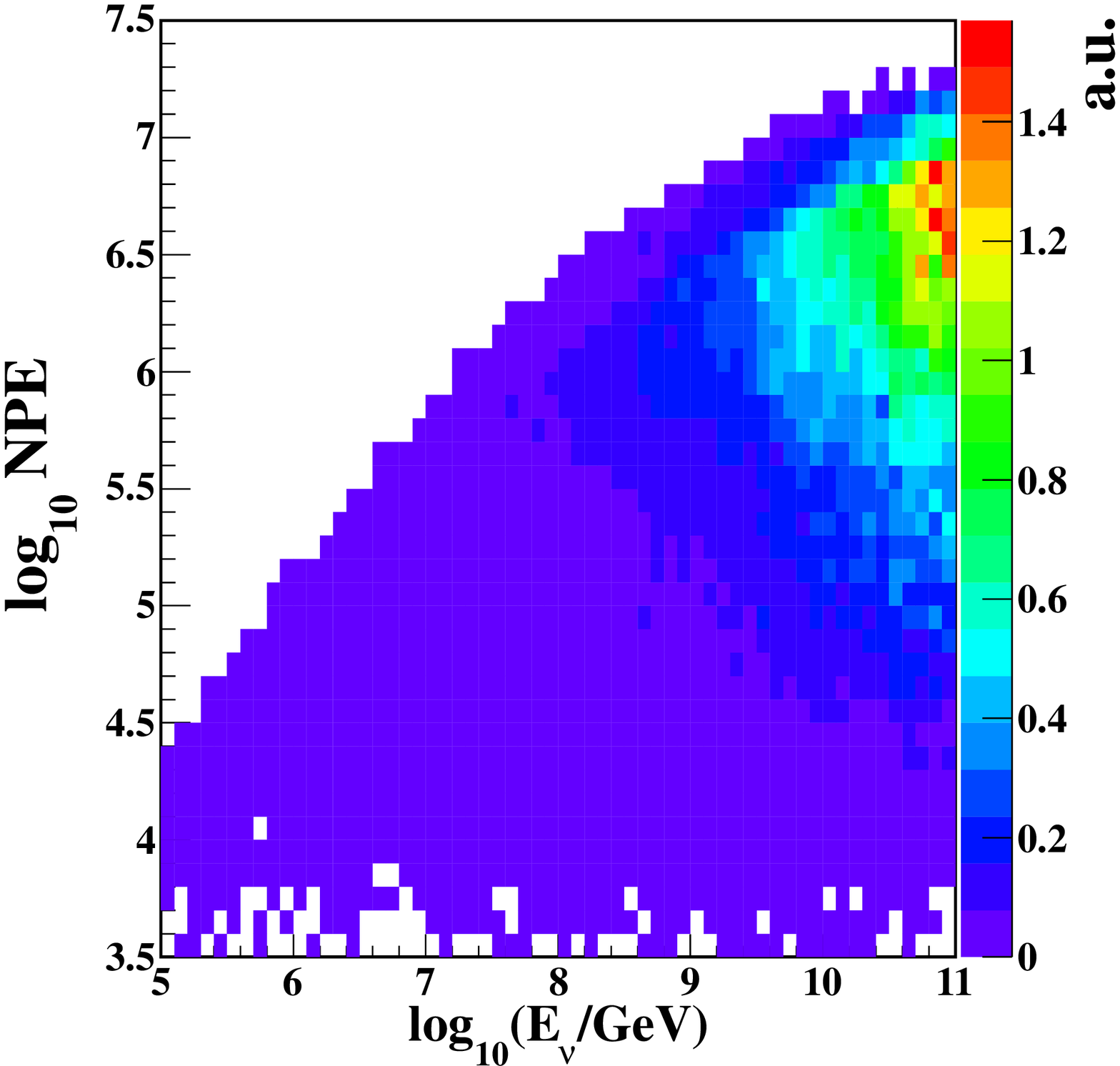}
  \caption{(color online). 
    Distributions of NPE versus the energies of neutrino-induced 
    muons (left) 
    and neutrinos which induce cascades (right)  obtained at the analysis 
    level with the IC86 signal Monte Carlo simulations.
    For illustrative purposes, an $E^{-1}$ energy spectrum of the particles 
    is assumed in these plots.
    The muon and neutrino energies are given when the particle enters
    a radius of 880~m around the IceCube center. Cascade events
    include all flavor neutral current and $\nu_e$ charged current 
    interactions.} \label{fig:energyNPE}
\end{figure}
%
\begin{figure*}[tbp]
  \includegraphics[width=1.85in]{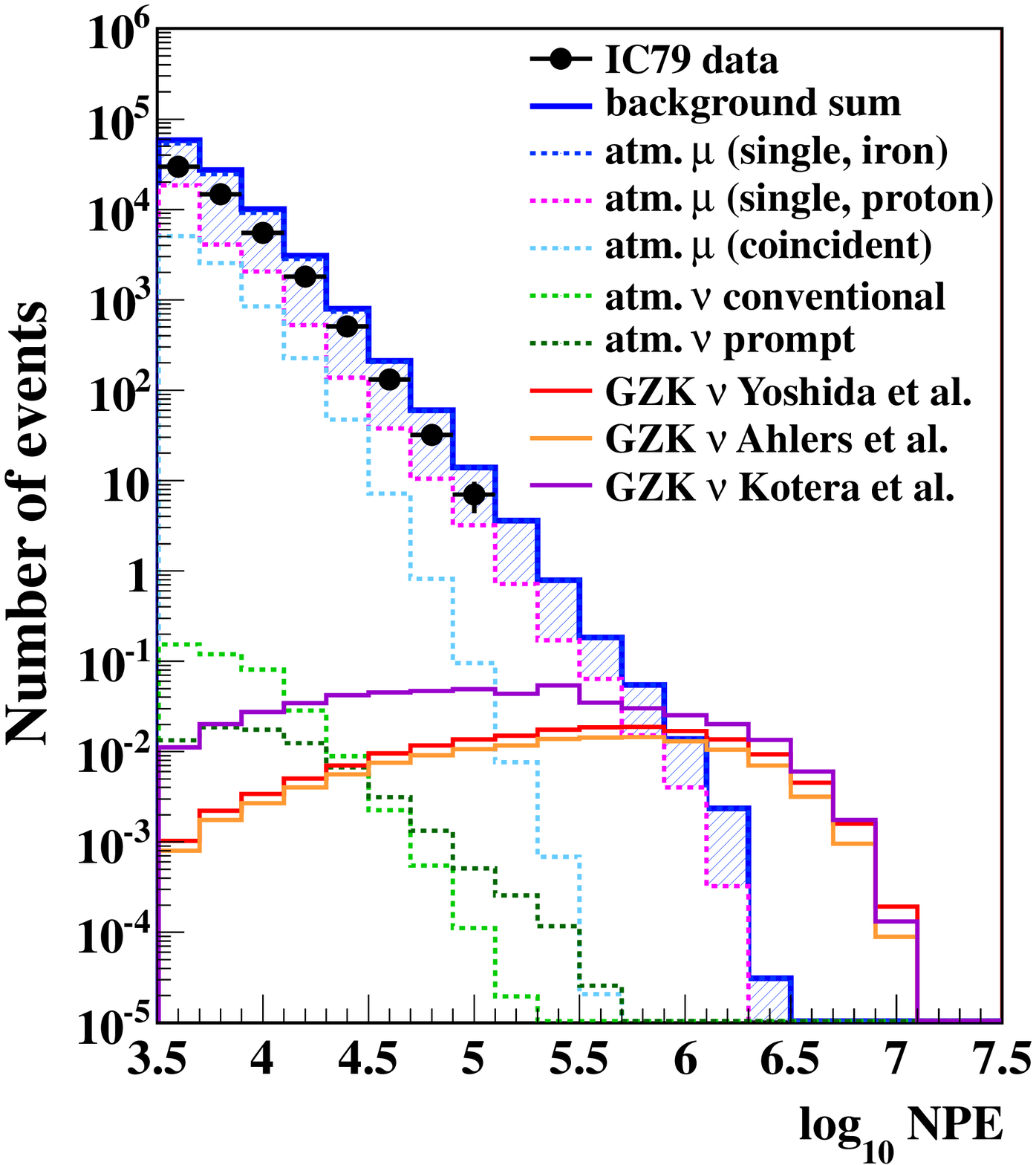}
  \includegraphics[width=1.85in]{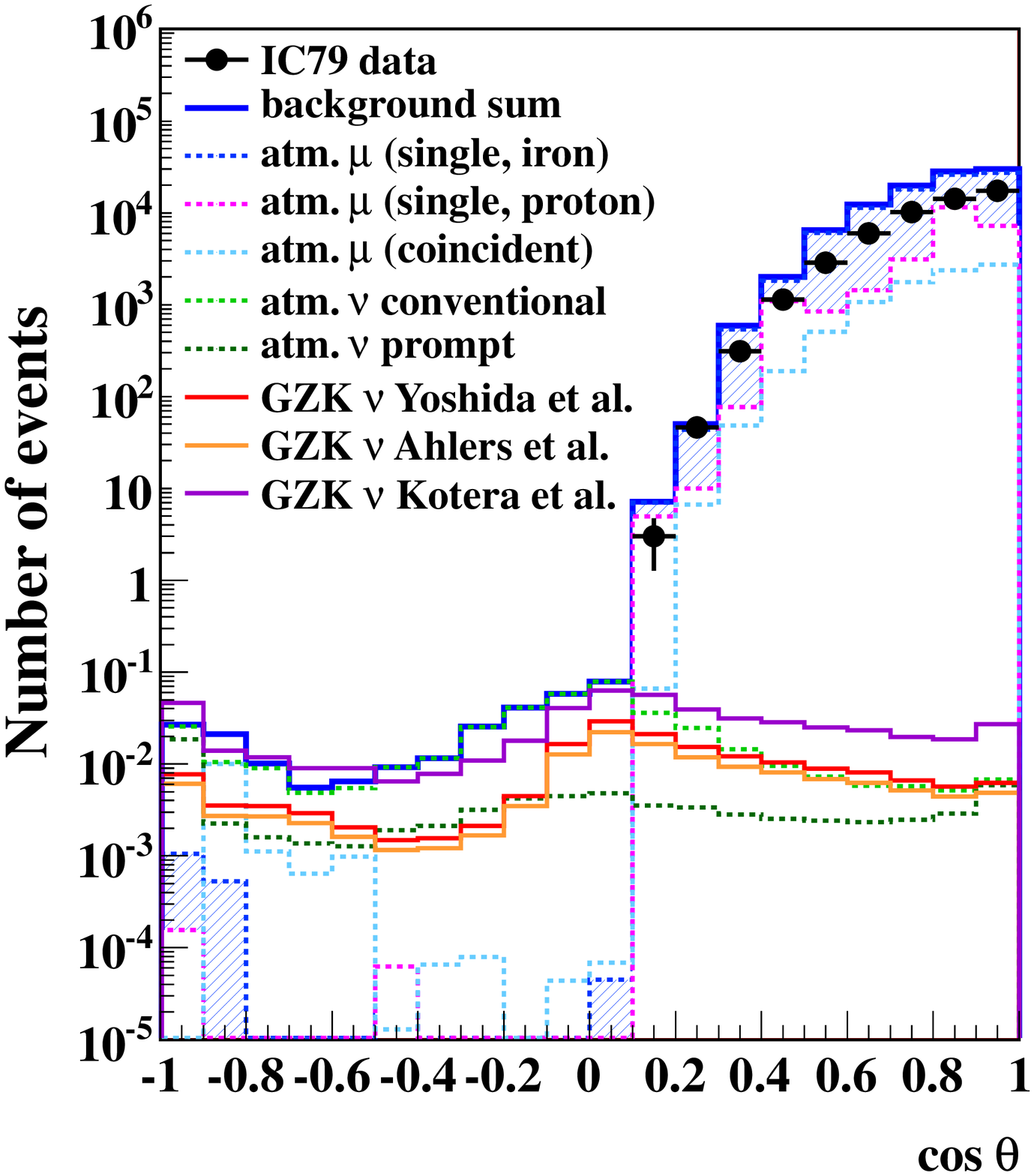}\\
  \includegraphics[width=1.85in]{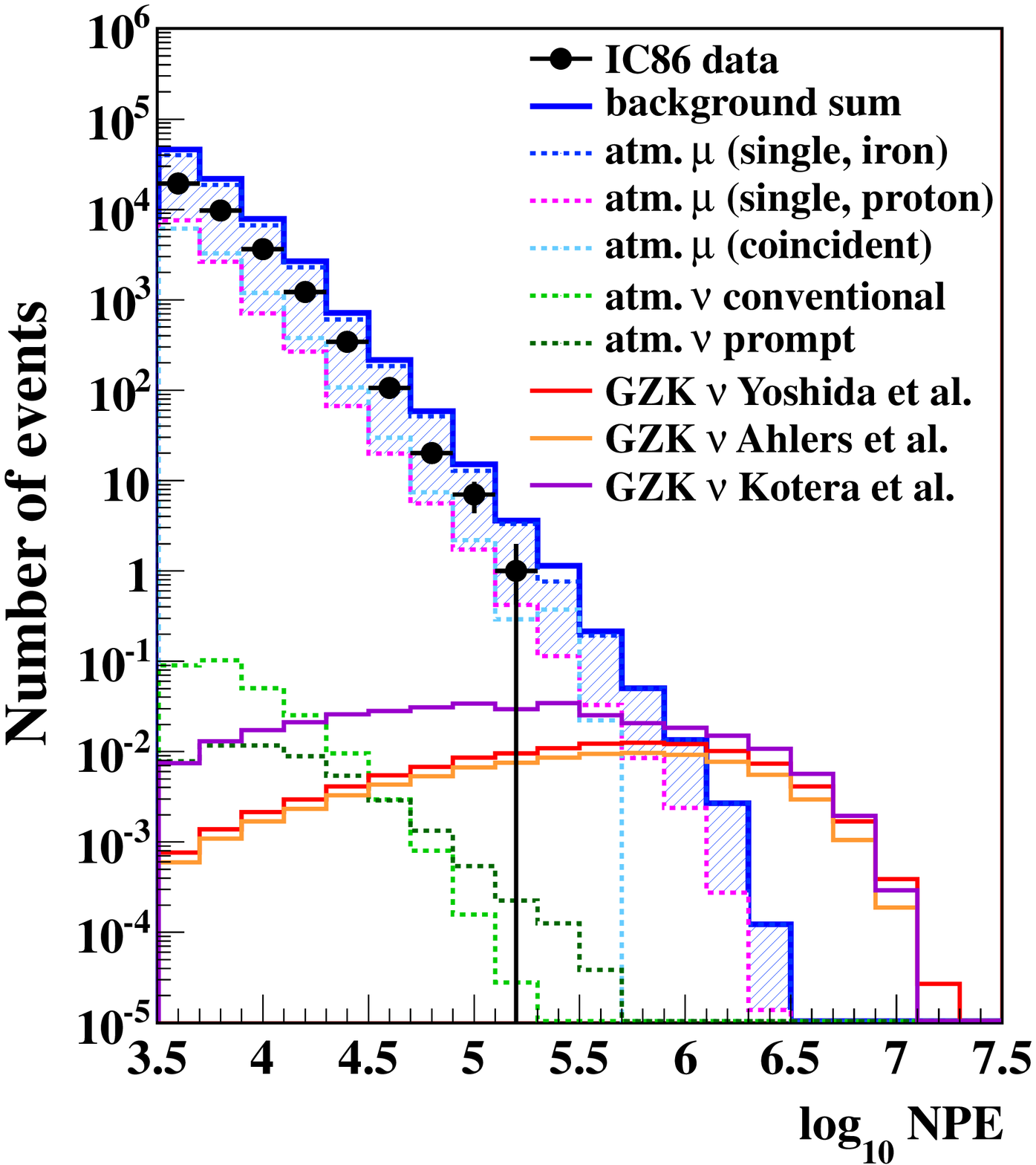}
  \includegraphics[width=1.85in]{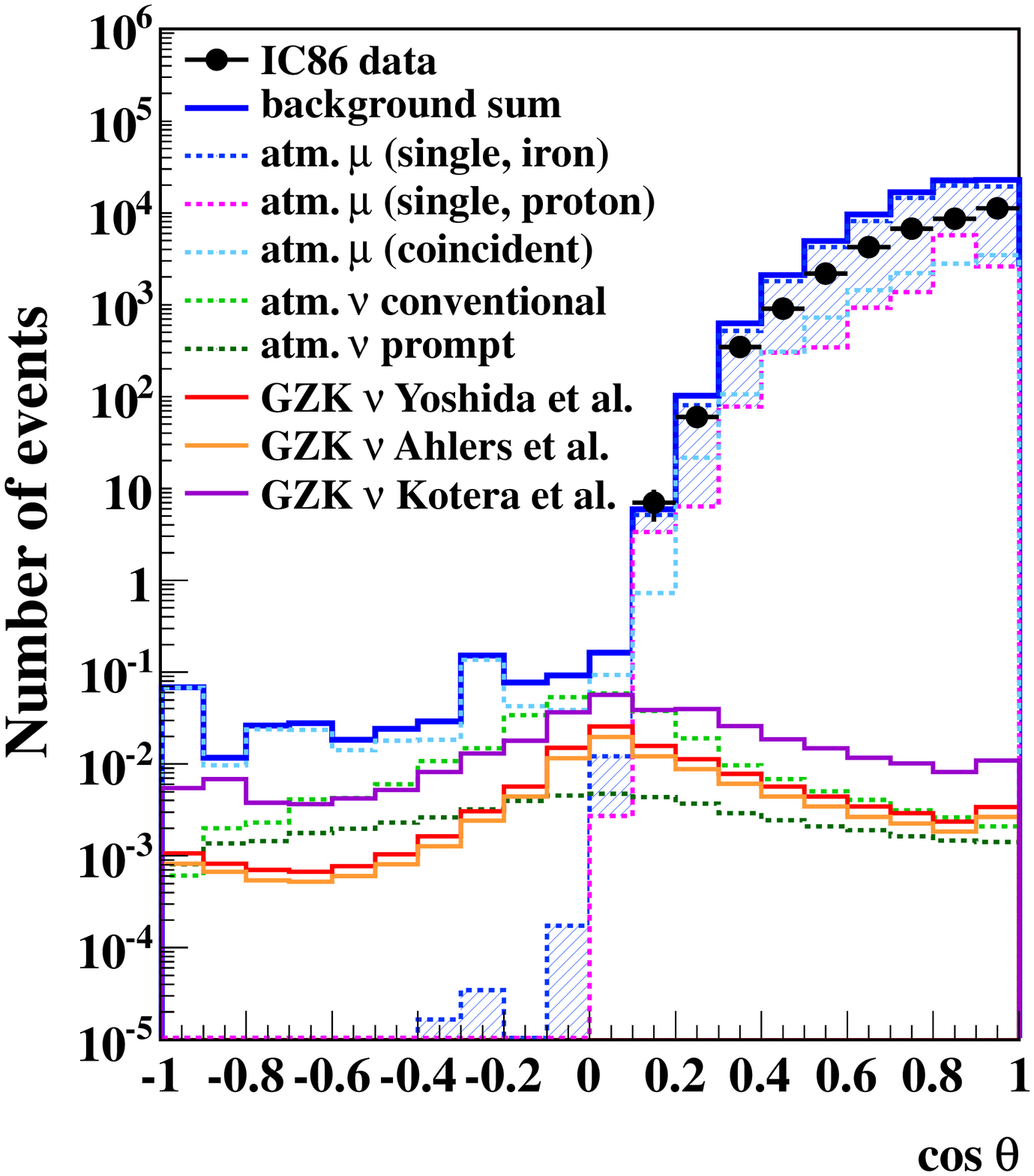}
  \caption{(color online). 
    Distributions of NPE (left panels) and reconstructed zenith angle 
    (right panels) are shown
    for the experimental test samples of IC79 (upper panels) and IC86 
    (lower panels).
    The data are compared with expected background contributions from 
    atmospheric muons and neutrinos, 
    and signals from various cosmogenic (GZK) neutrino 
    models~\cite{yoshida93, ahlers2010, kotera2010}.
    The event numbers presented here are for the livetimes of the test 
    samples of the experimental data, 
    33.4 days for IC79 and 20.8 days for IC86.
    The signal distributions are the sum of all three neutrino flavors. 
    The background sum includes all atmospheric muons and neutrinos. 
    The single atmospheric muons (pure iron) dominate the background 
    so that the line is nearly identical to the line for the background sum.
    See text for more detail.}
  \label{fig:analysislevel}
\end{figure*}
The distribution of total number of p.e.\ (NPE) versus the true energy of the incoming particle for 
IC86 simulations of neutrino-induced muons and cascades is shown in 
Fig.~\ref{fig:energyNPE}.
The energies are sampled when the incoming particle is at 880~m from the 
IceCube center.
A clear correlation between NPE and the energy of the muons is observed.
By selecting events with NPE above an appropriate threshold, 
low energy events, dominated by atmospheric backgrounds, are filtered out.
The correlation also holds for cascade events although uncontained events
with vertex positions outside the instrumentation volume
weaken the correlation
thereby reducing the selection efficiency for this type of event.

The left panels in Fig.~\ref{fig:analysislevel} show the NPE distributions at 
analysis level for data and simulations for IC79 and IC86, respectively. 
The signal cosmogenic neutrino distributions dominate over the atmospheric 
$\mu$ and $\nu$ background distributions in the high NPE region.
Three cosmogenic neutrino models are shown in the figure: 
Yoshida {\it et al.}~\cite{yoshida93} for an UHECR source distribution in the 
form of $(1+z)^m$ with the evolution parameter $m=4$ and the maximum redshift 
of the UHECR source distribution $z_{\rm max}=4$, 
Ahlers {\it et al.}~\cite{ahlers2010} (the best fit model with 
$m=4.6$ and $z_{\rm max}=2.0$)
and Kotera {\it et al.}~\cite{kotera2010} (Fanaroff-Riley type II).
Atmospheric muon bundles are the dominant contribution at this level.
Due to the yet unknown chemical composition of UHECRs, the background rates 
are estimated
by the extreme assumptions of pure proton and iron.
The pure iron is employed in this analysis as our baseline model
for the atmospheric muons since it yields more muons compared to the pure 
proton case and hence gives us a conservative background estimate.
For the pure iron case the predicted rate is about a factor of two higher 
than the rate observed in IceCube.
The data is bracketed by the two compositions as shown in 
Fig.~\ref{fig:analysislevel} by the shaded area,
demonstrating a reasonable agreement between the experimental data and the 
atmospheric muon background simulations. 

The directional information is also used to further discriminate signal from 
background.
Since the background of atmospheric muons is overwhelmingly large
compared to our signals above the horizon, a robust directional 
reconstruction is crucial for the discrimination. 
For this purpose, a track hypothesis is assumed to reconstruct atmospheric 
muons.
We utilized different zenith angle reconstruction algorithms for IC79 and 
IC86. 
A so-called single photo-electron (SPE) Log-Likelihood (LLH) fitting based 
on a track hypothesis using the probability distribution of the 
arrival time of the first photon in each DOM~\cite{Ahrens:2003fg} is 
performed for the IC79 sample. 
Then a cut on the reduced log-likelihood (rLLH) parameter is applied to 
ensure good fit quality.
The parameter rLLH is the log likelihood value of the reconstructed track 
divided by the number of degrees of freedom of the fit.
This rLLH cut removes coincident atmospheric muons.
For the IC86 sample, photon hits that have a significantly different timing 
compared to the one from the main bulk of photon signals are masked using 
the robust regression technique \cite{ilf}. 
Then the particle directions are reconstructed by applying the LineFit 
algorithm~\cite{icecubeEHE2010} to the remaining unmasked hits. 
The LineFit algorithm is based on a track hypothesis and uses a simple 
minimization of 
$\chi^2 = \Sigma_i {\rm NPE}_i (\vec{r_i}-\vec{r}_{\rm COG}-t_i\vec{v})^2$, 
where $t_i$ and ${\rm NPE}_i$ represent the time of the first photo-electron 
and the number of photo-electrons recorded by the {\it i}th DOM at the 
position $\vec{r_i}$,  respectively. 
The quantity $\vec{r}_{\rm COG} \equiv (\frac{\Sigma_i {\rm NPE}_i \;
{x_i}}{\Sigma_i {\rm NPE}_i}, \frac{\Sigma_i {\rm NPE}_i \;
{y_i}}{\Sigma_i {\rm NPE}_i}, \frac{\Sigma_i {\rm NPE}_i \;
{z_i}}{\Sigma_i {\rm NPE}_i})$ 
is the position of the NPE-weighted center-of-gravity of the hits. 
The fit ignores the geometry of the Cherenkov cone and the optical properties 
of the medium and assumes light traveling with a velocity $\vec{v}$ along a 
1-dimensional path through the detector, passing through the center-of-gravity.
The inclusion of the robust regression technique significantly 
improves the performance of the LineFit used in the previous 
study~\cite{icecubeEHE2011}, allowing for simpler background rejection.
The zenith angle resolution of SPE LLH for background muon events is 
about $0.5^\circ$ for the IC79 EHE analysis level sample.
The zenith angle resolution from the LineFit with the robust regression for 
background muons for the IC86 analysis level sample
is about $1^{\circ}$. These performances are sufficient to remove atmospheric 
muon bundle background events in the current analysis.

The performance of the reconstruction on the signal neutrinos highly depends 
on the shape of the events 
(Fig.~\ref{fig:eventviews}).
Since most of the signal neutrino events ($> 80\%$) are expected to be muon 
or tau tracks,
the reconstruction of zenith angles based upon track hypotheses as described 
above gives sufficiently good signal selection efficiency. 
The reconstructed directions of neutrino induced cascades, however,
are only poorly correlated with the true neutrino direction and exhibit 
systematic directional shifts.
The SPE LLH reconstruction tends to shift the zenith angles 
towards the vertical while the LineFit shifts them to the horizontal.
The behavior of the shifts also changes when their vertex positions are 
close to or outside the boundary of the instrumentation volume.
The resulting systematic uncertainty is discussed in Section~\ref{sec:Sys}.

\begin{table*}[tbp]                                                         
\caption{
  Number of events passing cuts at on-line filtering, off-line analysis, and 
  final level with
  285.8~days of effective livetime for IC79 and 330.1~days for IC86 
  (excluding test sample data). 
  One cosmogenic neutrino model~\cite{yoshida93} (with $m=4$ 
  and $z_{\rm max}=4$) 
  is taken to evaluate the benchmark signal rates.
  The background rates include atmospheric muons assuming a pure iron 
  primary composition,
  conventional atmospheric neutrinos, and prompt atmospheric neutrinos.
  Analysis sample requests the number of hit DOMs $\geq$ 300, 
  log$_{10}$ (NPE) $\geq$ 3.5 for IC79 and IC86, and an additional 
  requirement of rLLH $<$ 8 for IC79.
  Systematic uncertainties in the expected
  event rates at the final selection level are given as asymmetric error
  intervals after the statistical errors.
  \label{table:level2}}
\begin{ruledtabular}
  \begin{tabular}{lrrrrrr}
    Contributions
      &  \multicolumn{2}{c}{Experimental}
      &  \multicolumn{2}{c}{Background MC}
      &  \multicolumn{2}{c}{Benchmark signal MC \cite{yoshida93}} \\
   {Samples}
   & IC79 & IC86 & IC79 & IC86 & IC79 & IC86 \\
   \hline
   &  &  &  &  &  &  \\
   EHE filter level
   & $4.0\times10^7$ & $6.0\times10^7$  & $4.4\times10^7$ &  $8.9\times10^7$  
   & 2.1  & 2.4  \\
   Analysis level & $4.5\times10^5$ & $5.9\times10^5$  & $8.5 \times 10^5$ 
   &  $1.3 \times10^6$  & 1.5 & 1.8 \\
   Final level &  0  &  2  &  0.056$\pm$0.002$^{+0.028}_{-0.041}$ 
   &  0.026$\pm$0.003$^{+0.015}_{-0.017}$
   & 0.876$\pm$0.004$^{+0.119}_{-0.105}$& 1.043$\pm$0.006$^{+0.142}_{-0.134}$\\
  \end{tabular}
 \end{ruledtabular}
\end{table*}

The right panels in Fig.~\ref{fig:analysislevel} show the event distributions 
at analysis level 
as a function of the cosine of the reconstructed zenith angle. 
These distributions are compared to the background and signal simulations. 
Atmospheric muon bundles dominate in the downward-going region and 
atmospheric neutrinos dominate in the upward-going region.

\begin{figure*}[btp]
\begin{center}
  \includegraphics[width=1.7in]{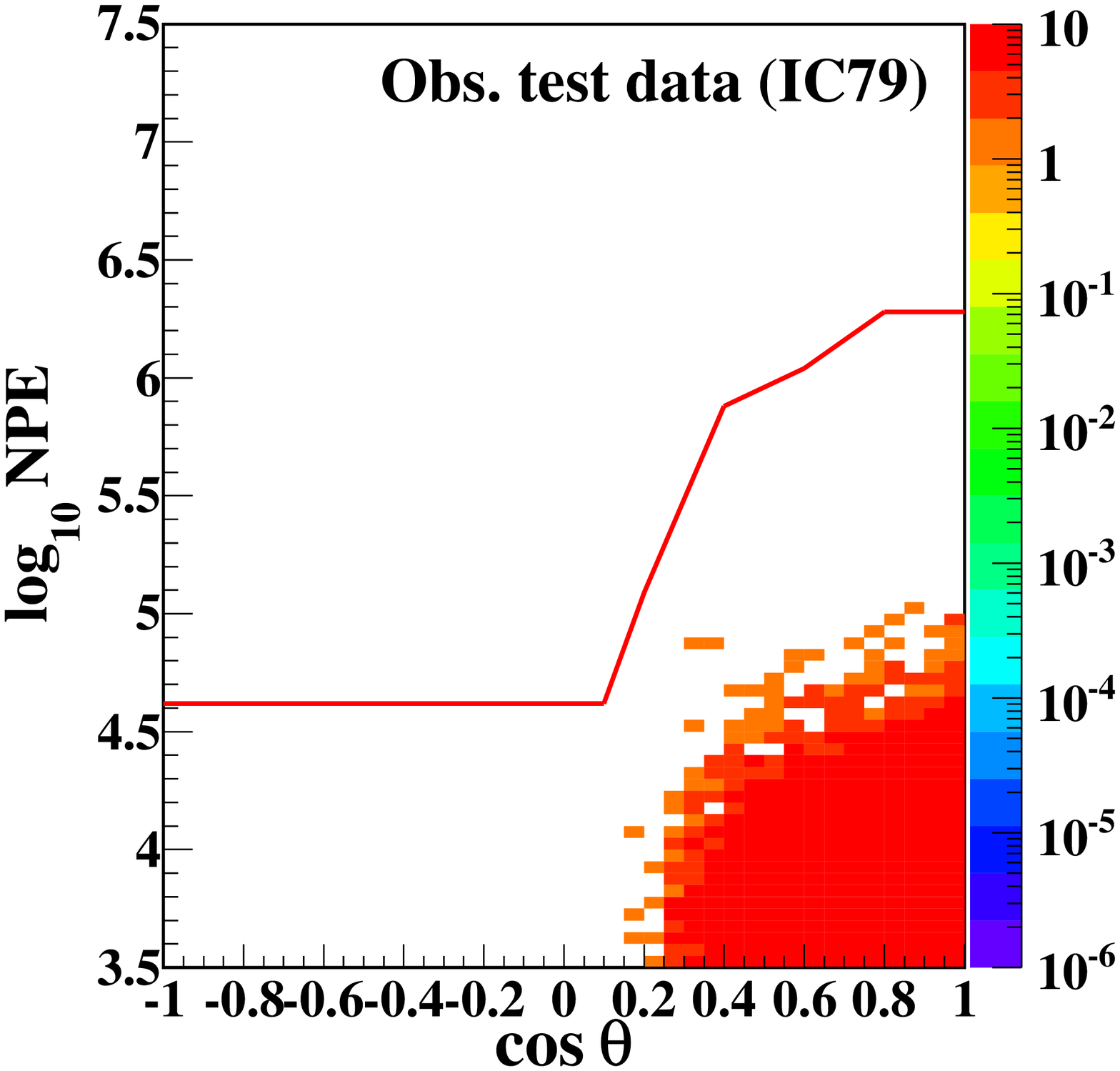}
  \includegraphics[width=1.7in]{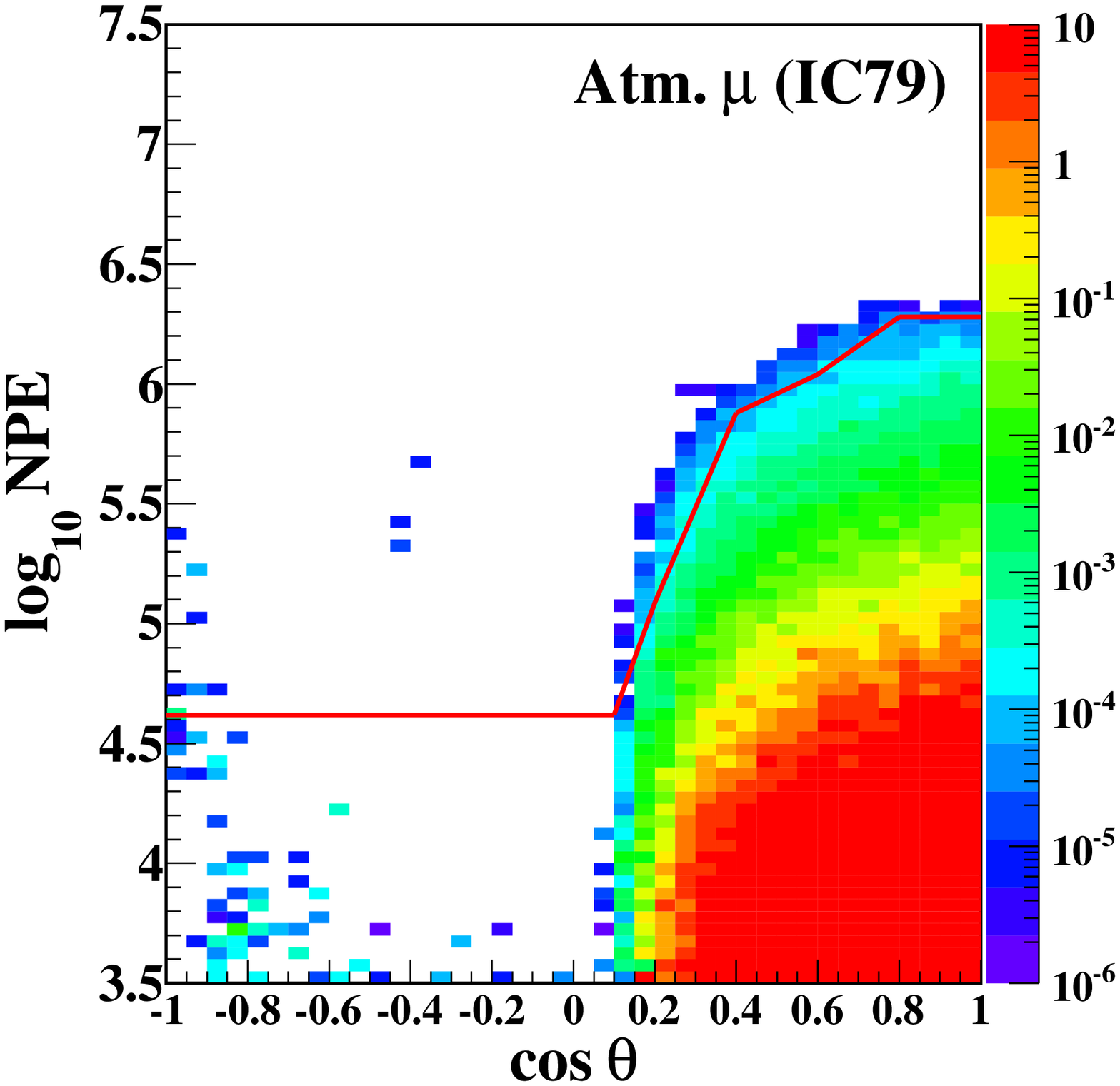}
  \includegraphics[width=1.7in]{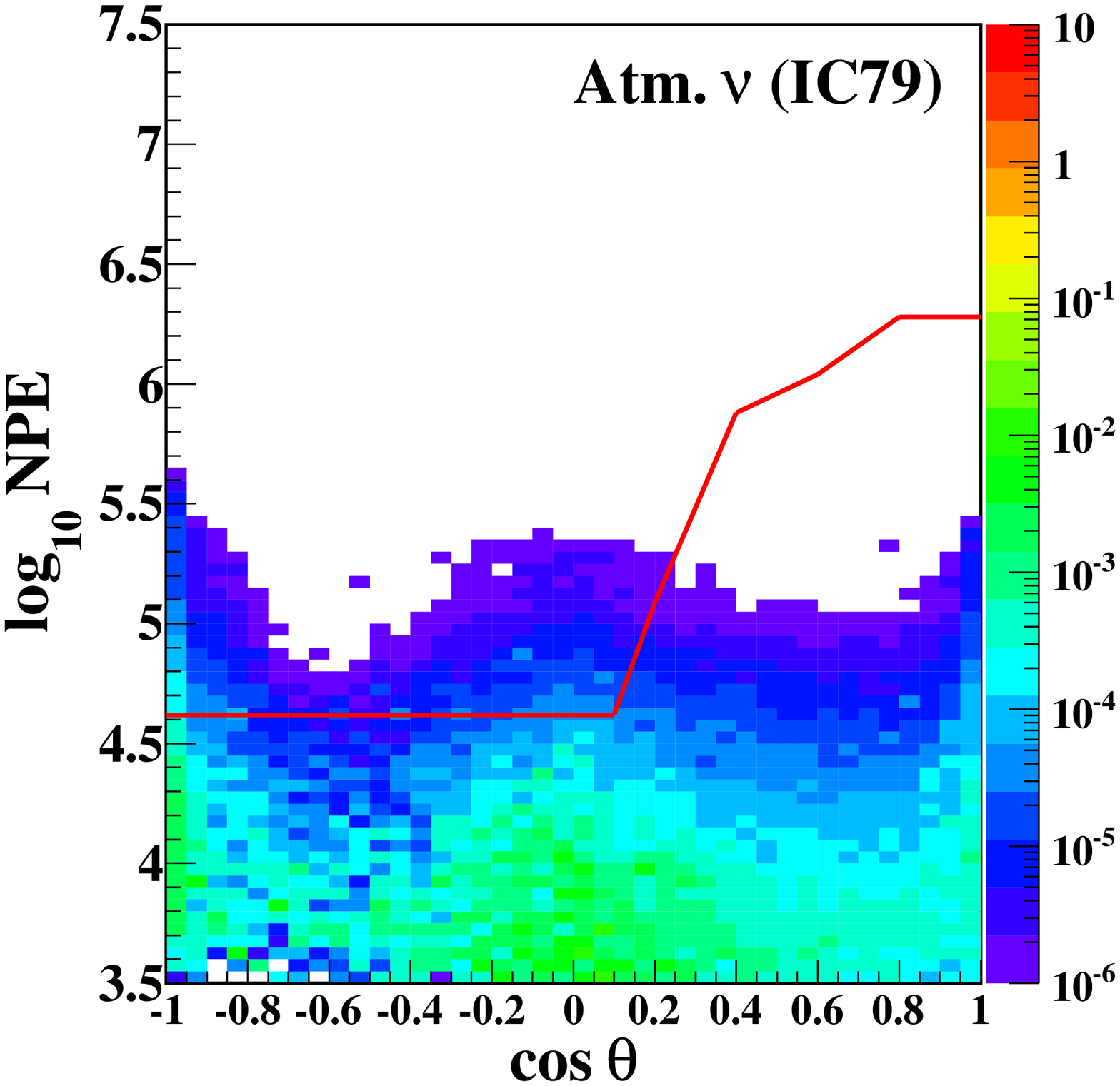} 
  \includegraphics[width=1.7in]{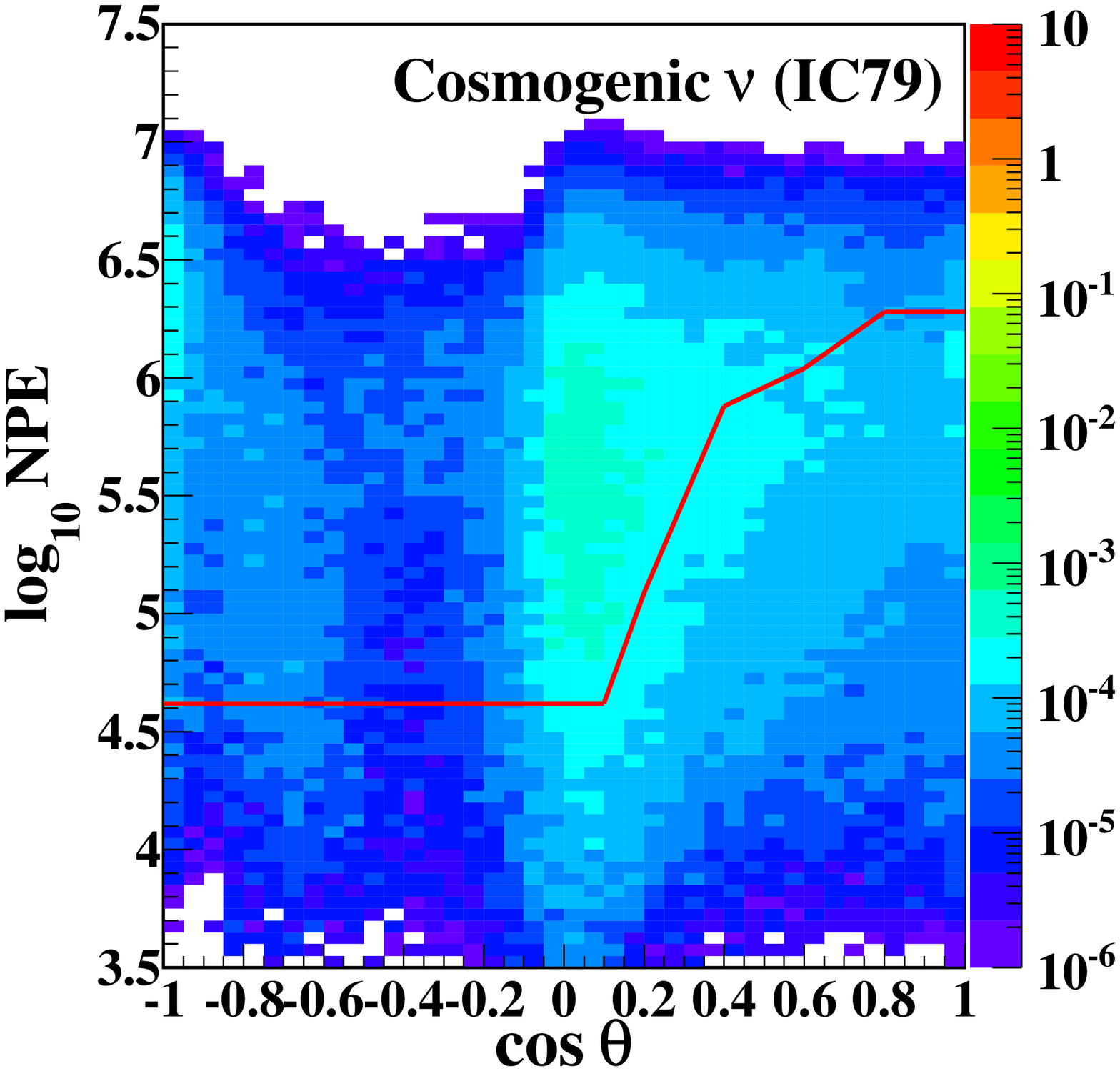}\\
  \includegraphics[width=1.7in]{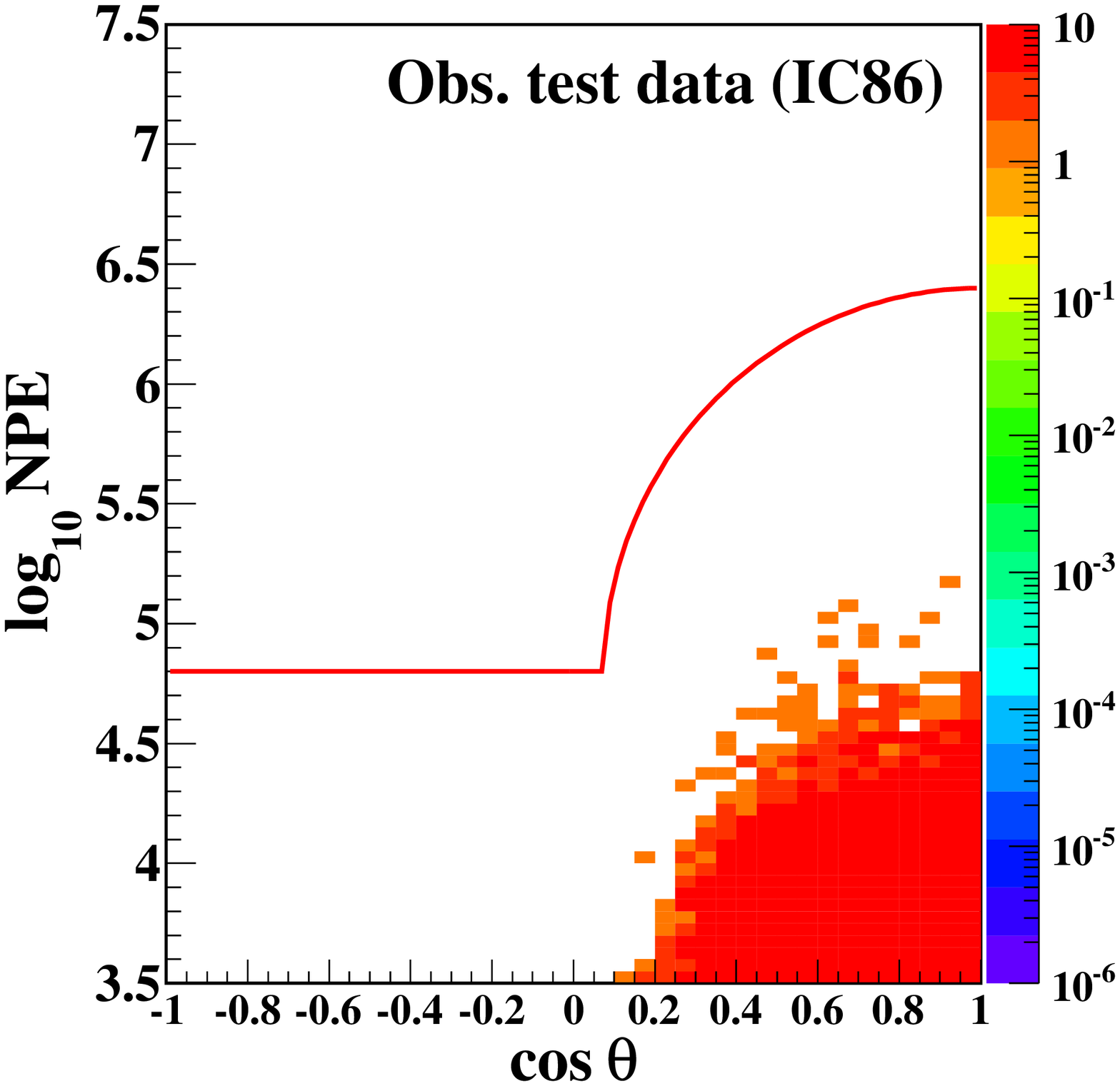}
  \includegraphics[width=1.7in]{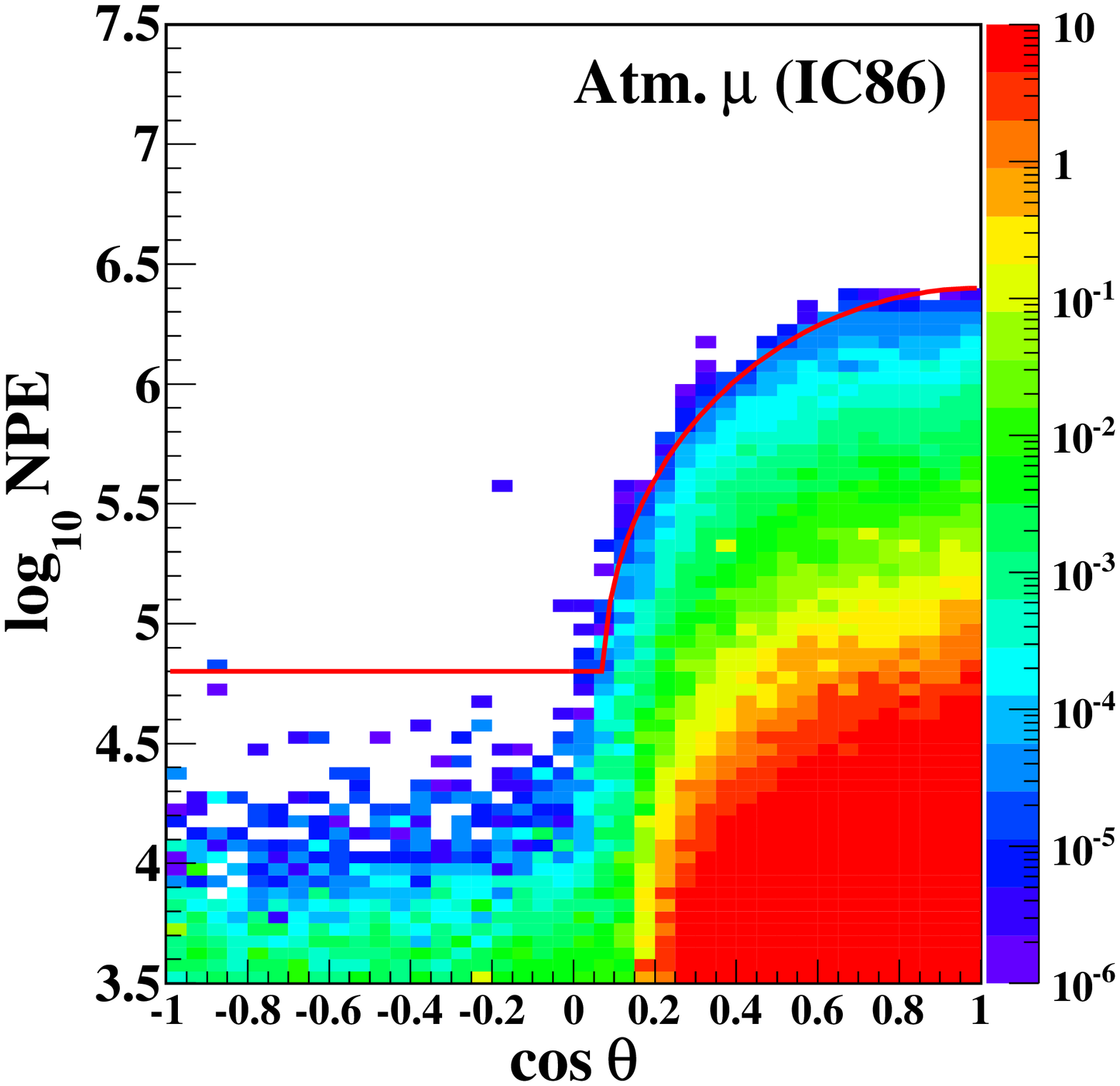}
  \includegraphics[width=1.7in]{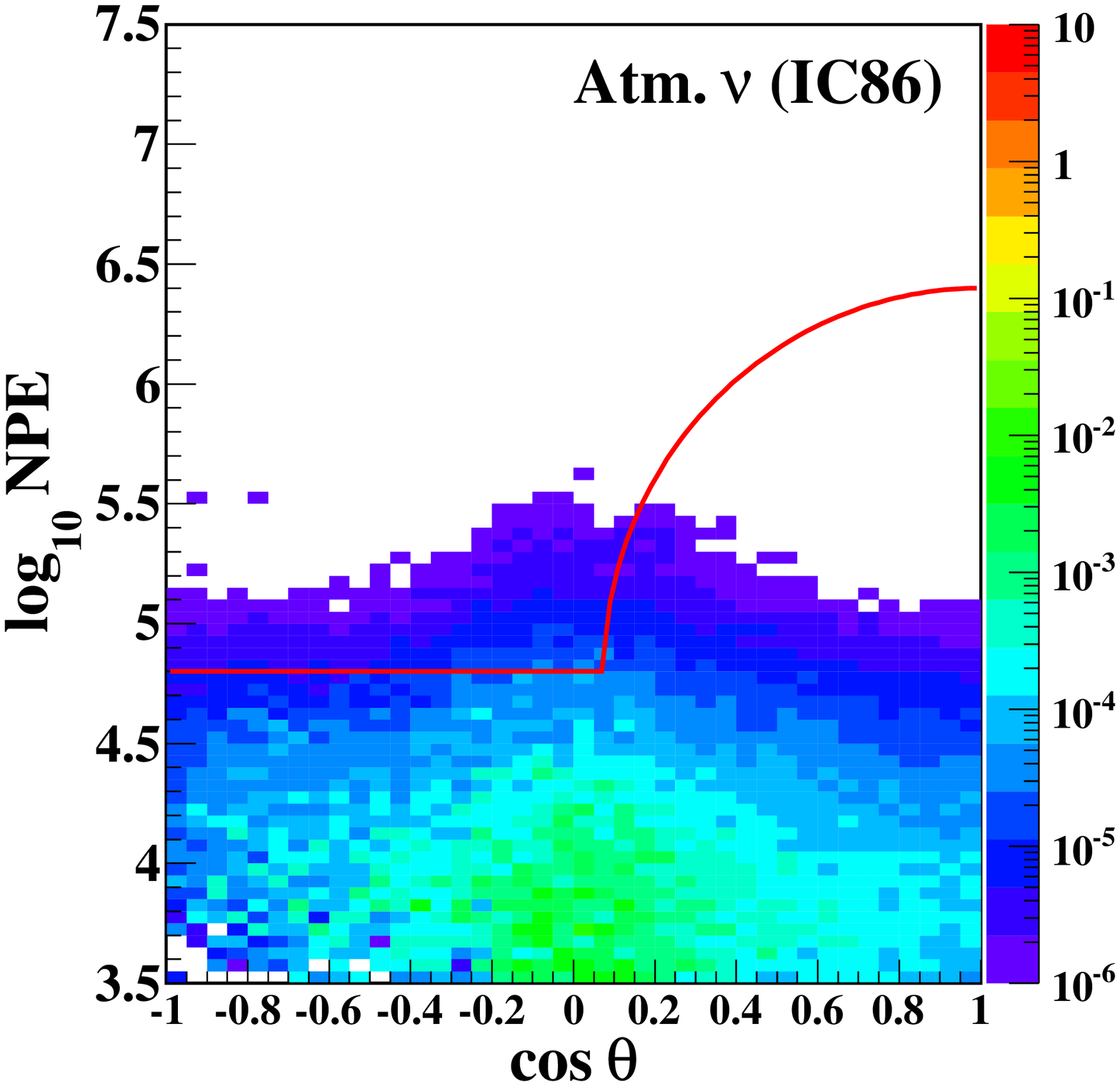}
  \includegraphics[width=1.7in]{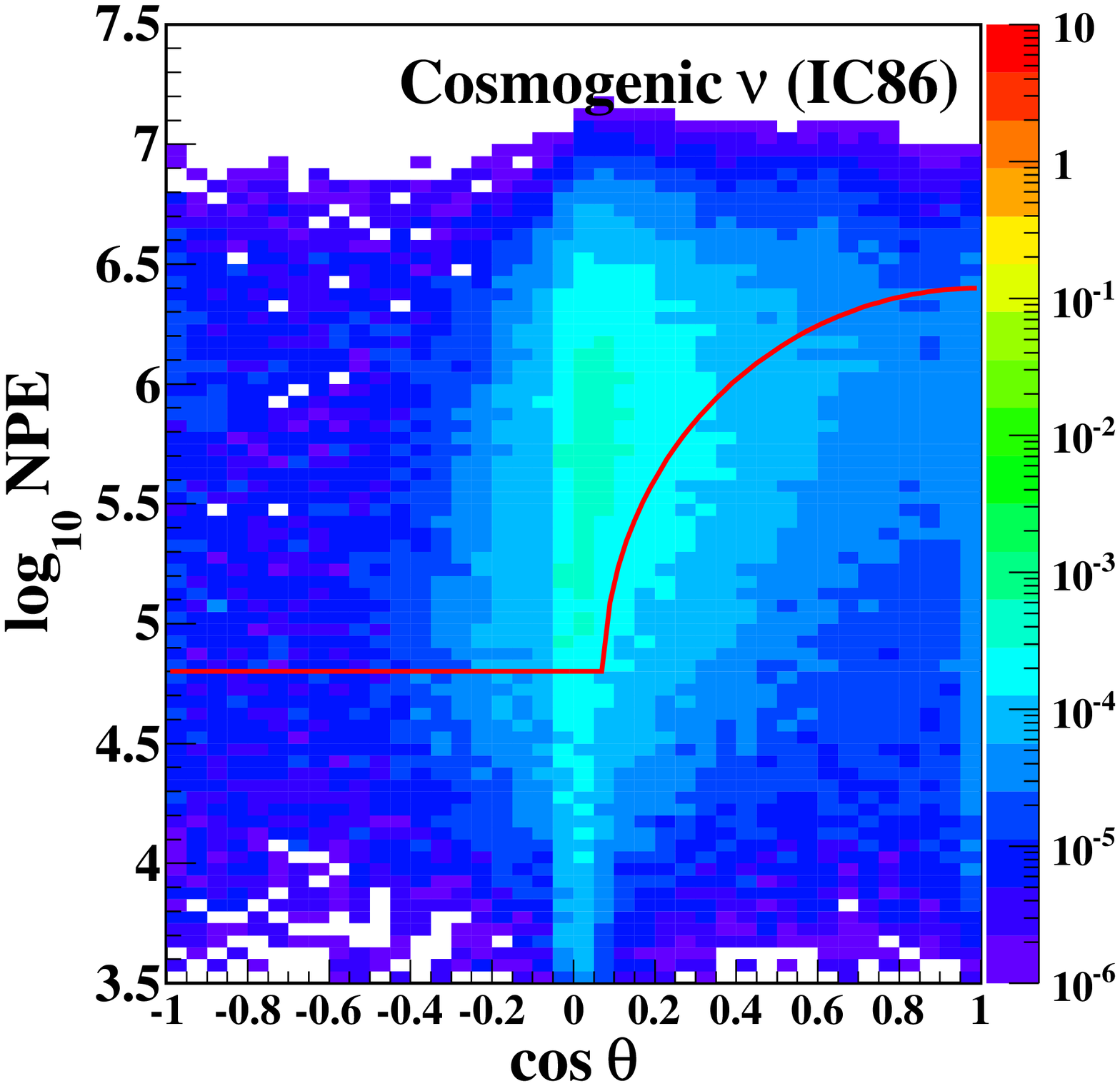}
\end{center}
\caption{(color online).
  Event number distributions on the plane of NPE and cosine of 
  reconstructed zenith angle ($\cos\theta$)
  for the IC79 run (upper panels) and the IC86 run (lower panels).  
  The experimental test samples are shown in left panels. 
  The background simulations of atmospheric muon (middle left panels), 
  and the conventional atmospheric neutrino and prompt atmospheric 
  neutrino~\cite{prompt_enberg} (middle right panels), 
  and simulation of signal cosmogenic neutrino model~\cite{yoshida93} 
  (right panels) are also shown. The colors indicate event numbers per 
  livetime of 33.4~days and 20.8~days for the IC79 and IC86 test samples 
  respectively. The signal distributions are the sum of all three neutrino 
  flavors. 
  The solid lines in each panel indicate the final selection criteria.}
\label{fig:final}
\end{figure*}

\begin{figure*}[tbp]
  \begin{center}
    \includegraphics[height=2.1in, width=2.2in]{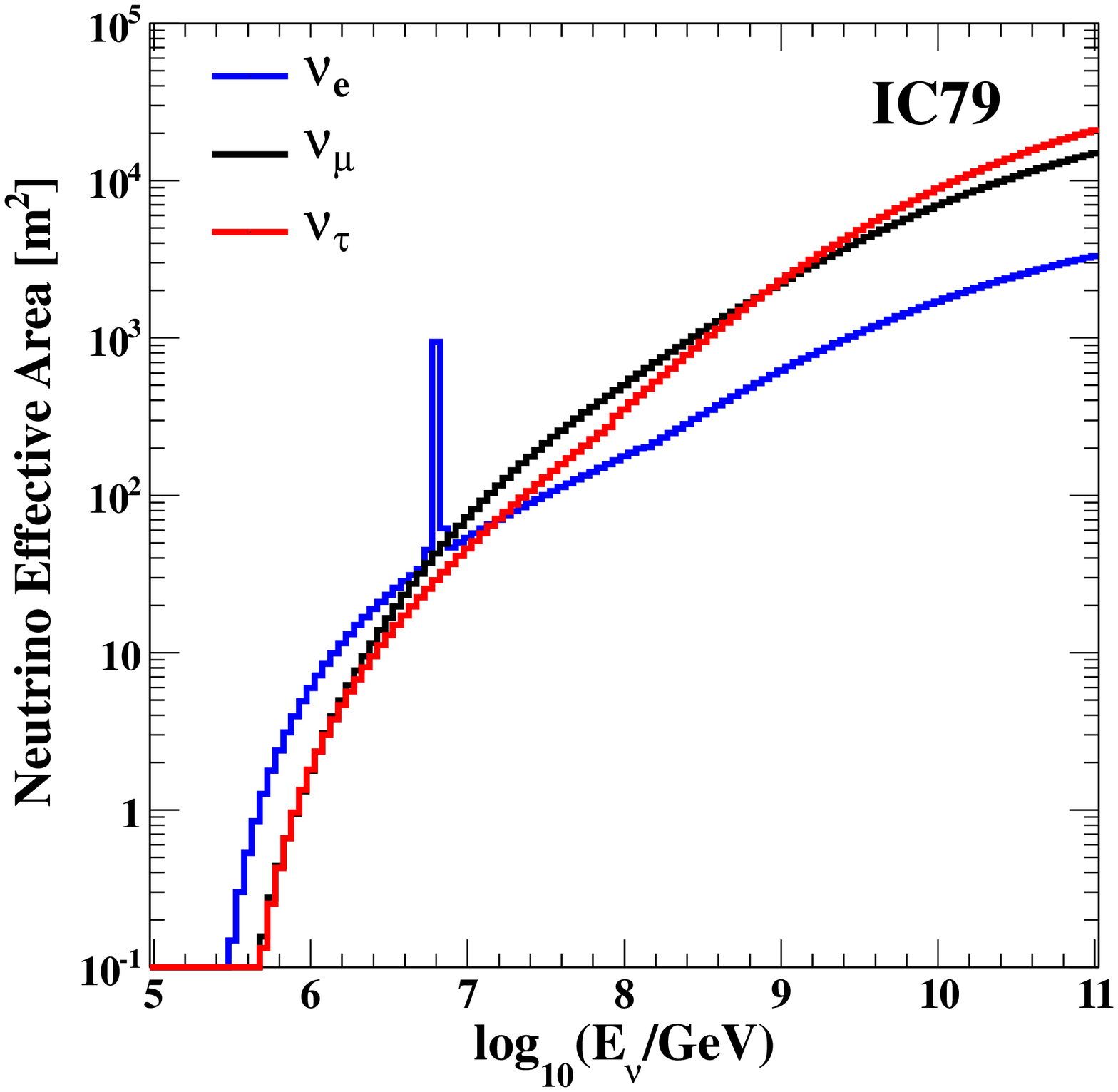}
    \includegraphics[height=2.1in, width=2.2in]{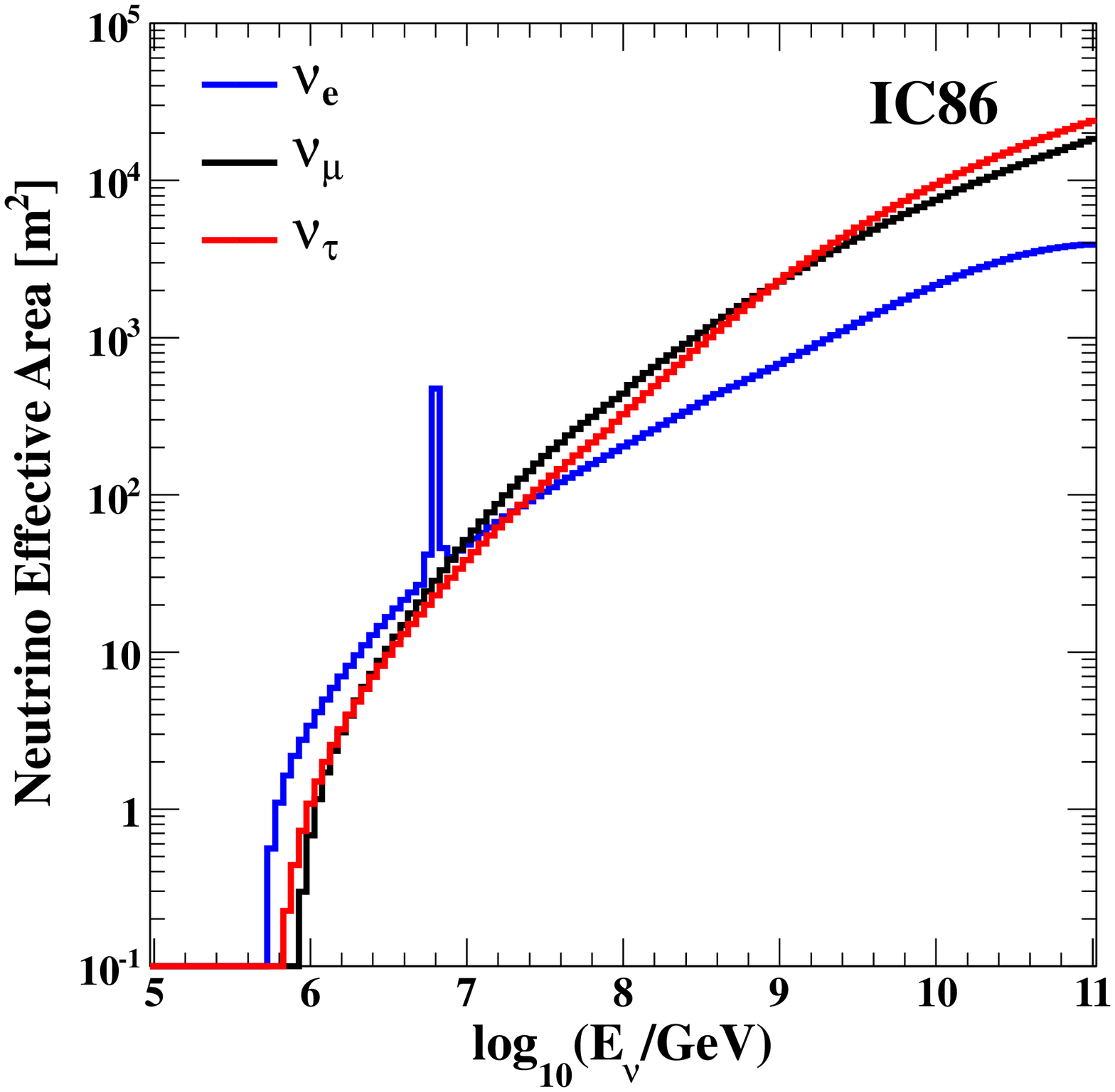}
    \includegraphics[height=2.1in, width=2.2in]{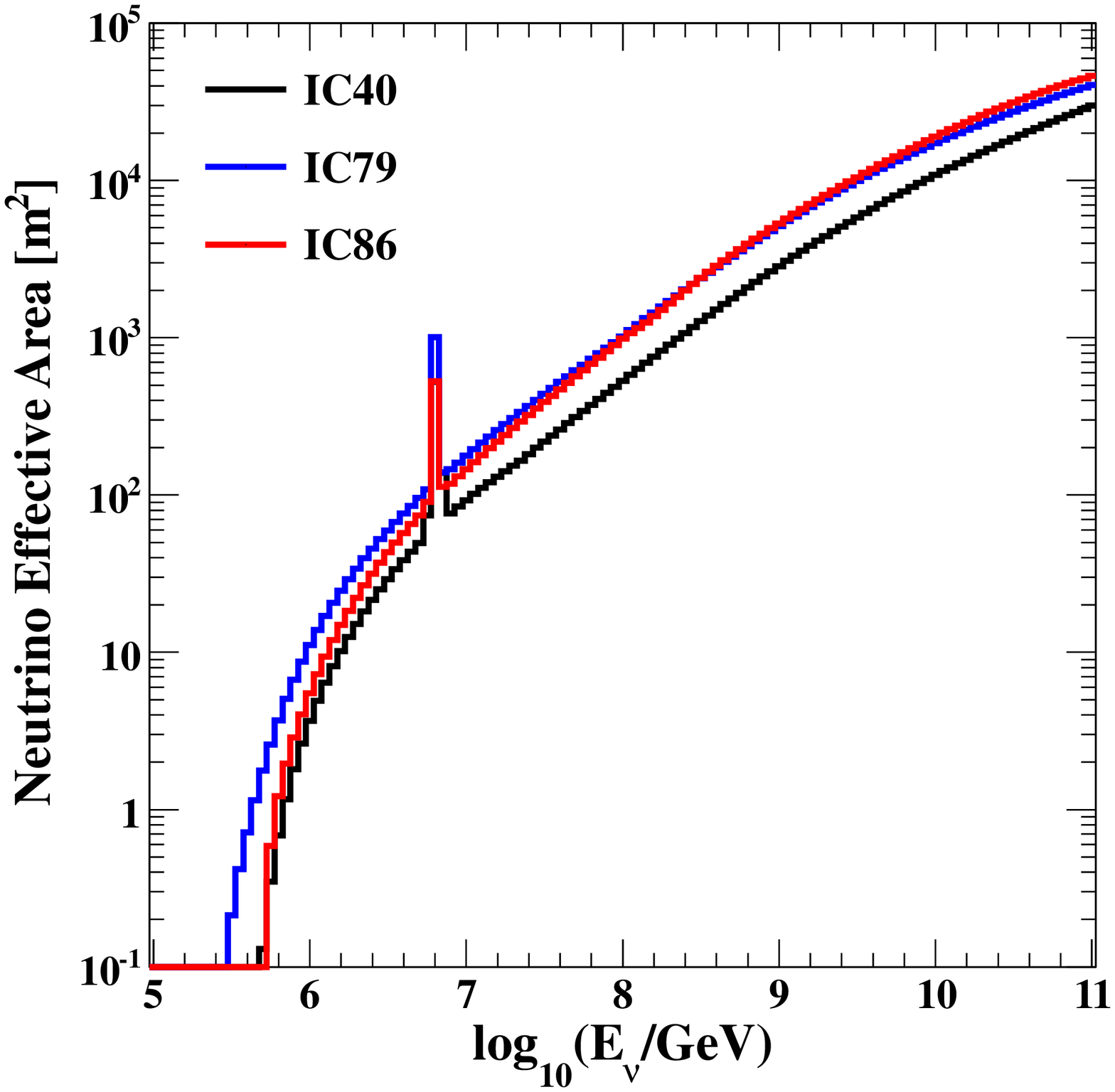}
  \end{center}
  \caption{(color online).
    The IceCube neutrino effective area at final selection criteria 
    with different string configurations, IC79 (left panel) and IC86 
    (middle panel) for each neutrino flavor, averaged over 4$\pi$ solid angle.
    The areas are averaged over equal amounts of neutrinos and anti-neutrinos. 
    Three flavor sums of the effective areas are shown in the right panel. 
    The effective area from the previous search~\cite{icecubeEHE2011} with 
    40 string configuration of IceCube (IC40) is also shown for comparison. 
    Exposure of the sample used in this analysis is obtained by multiplying
    the effective area with the effective livetime without test samples 
    (333.5~days, 285.8~days and 330.1~days for IC40, IC79, and IC86 
    respectively) and 4$\pi$ solid angle. The sharp peaked structure at 
    6.3~PeV for electron neutrinos is due to the Glashow 
    resonance~\cite{glashow}.
}
  \label{fig:effarea}
\end{figure*}

The signal selection criteria were optimized based on simulations of 
background and signal 
after the simulation was verified using the test sample.
A cosmogenic neutrino model~\cite{yoshida93} (with $m=4$ 
and $z_{\rm max} = 4$) is used for the optimization.
The selection criteria do not severely depend on the particular choice of 
the cosmogenic model since the expected energy spectrum is similar.
Selection criteria are obtained by optimizing the NPE threshold values in 
the IC79 and IC86 samples separately such that the model discovery 
factor~\cite{MDF, icecubeEHE2011} is minimized in each sample.
Fig.~\ref{fig:final} presents the event distributions in the plane of 
NPE vs cosine of the reconstructed zenith angle ($\cos\theta$) for the 
test sample and simulations.
The distributions of the signal simulation are the sum of all three 
neutrino flavors. The solid lines in Fig.~\ref{fig:final} indicate the 
final selection criteria for each sample. The events above the lines are 
considered to be signal event candidates. 
The essential point of this analysis is to select high NPE events against
backgrounds regardless of the event shape.
A zenith angle dependent high NPE threshold is 
required to eliminate the atmospheric muon background for the downward-going 
region, while a constant threshold value is placed in the zenith region of 
cos $\theta$ $\lesssim$ 0.1, where no atmospheric muon background is expected.
The predicted number of signal and background events passing the final 
selection criteria are presented
in Table~\ref{table:level2} along with the observed number of events in the 
two experimental samples.

The effective neutrino detection areas at final selection criteria for the 
different IceCube detector configurations are shown in Fig.~\ref{fig:effarea}.
 The effective areas are given for each neutrino flavor, averaged over 
4$\pi$ solid angle for IC79 and IC86.
The areas are averaged over equal fluxes of neutrinos and anti-neutrinos. 
Below 5~PeV, the effective area for electron neutrinos exceeds that of muon 
or tau neutrinos. 
For particle cascades induced by charged current interactions of electron 
neutrinos, their energies are deposited completely inside the detector 
if their interaction vertex lies sufficiently inside the instrumented volume.
Contrarily muons (taus) from muon (tau) neutrino interactions only partially 
deposit their energies in the detector volume.
Therefore, even though tracks have a longer path in the detector, they 
satisfy the NPE criteria less frequently (Fig.~\ref{fig:final}).
At higher energies the effective area for tracks is larger
because they can be generated in an increasingly larger volume and still 
reach the detector. 
Above 100 PeV cascades contribute less than 20\% to the observable events 
from cosmogenic neutrino fluxes.
The right panel in Fig.~\ref{fig:effarea} shows the effective area summed 
over all three neutrino flavors for IC79 and IC86 together with that for 
IC40 from the previous analysis \cite{icecubeEHE2011}. The current analysis 
has approximately a factor of two larger effective area compared to IC40. 
The difference between the effective areas for IC79 and IC86 below 30\,PeV 
originates from the different NPE thresholds. The slight difference 
above $3 \times 10^{3}$\,PeV is due to the rLLH cut in IC79.


\section{\label{sec:Sys} Systematic uncertainties}

\begin{table*}[htb]
\begin{center}
\caption{List of the statistical and systematic errors on the signal, 
atmospheric muon and neutrino, prompt neutrino and the total background rate. 
The uncertainties in the signal rate are estimated for the cosmogenic flux 
of Yoshida {\it et al.}~\cite{yoshida93} for $(m,z_{\rm max})=(4,4)$. 
The uncertainties in the background rates are evaluated against the baseline 
estimation by CORSIKA-SIBYLL~\cite{Heck:1998vt,Ahn:2009wx}
with a pure iron composition hypothesis for atmospheric muons and the 
Gaisser-H3a model~\cite{TG} for atmospheric neutrinos. 
The uncertainties in the prompt neutrino rate are estimated using the 
prediction by Ref.~\cite{prompt_enberg}.
The systematic and statistical errors listed here are relative to the event 
rates for each signal and background source.}
\label{tb:Sys}

\begin{tabularx}{15.cm}{c c c c c c} \hline \hline
                                      &                      &
                      & Conventional       &                    &       \\
  Sources                             & Cosmogenic           & 
Atmospheric          & atmospheric        & Prompt             & Total      \\
                                      & $\nu$ signal (\%)    & 
muon (\%)            & neutrino (\%)      & neutrino (\%)      & 
background (\%)\\ \hline
  Statistical error                   & $\pm$0.4             & 
$\pm$9.1             & $\pm$9.8           & $\pm$1.1           & 
$\pm$4.5 \\ \hline
  DOM efficiency                      & $^{+1.5}_{-5.1}$     & 
$^{+41.9}_{-42.7}$   & $^{+73.2}_{-17.9}$ & $^{+33.6}_{-9.6}$  & 
$^{+43.1}_{-26.1}$ \rule[-2mm]{0mm}{6mm} \\
  Ice properties/Detector response    & $-$7.2               & 
$-$47.7              & $-$44.8            & $-$30.8            & $-$41.7 \\
  Neutrino cross section              & $\pm$9.0             & 
$-$                  & $-$                & $-$                & $-$ \\
  Photo-nuclear interaction           & $+$10.0              & 
$-$                  & $-$                & $-$                & $-$ \\
  LPM effect                          & $\pm$1.0             & 
$-$                  & $-$                & $-$                & $-$ \\ 
  Angular shift for cascades          & $-$0.5               & 
$-$                  & $-$                & $-$                & $-$ \\
  Cosmic-ray flux variation           & $-$                  & 
$^{+30.0}_{-50.0}$   & $\pm$30.0          & $\pm$30.0          & 
$^{+18.7}_{-26.3}$ \rule[-2mm]{0mm}{6mm} \\
  Cosmic-ray composition              & $-$                  & 
$-$79.1              & $-$                & $-$                & $-$36.7 \\
  Hadronic interaction model          & $-$                  & 
$+$17.7              & $-$                & $-$                & $+$8.1\\
  $\nu$ yield from cosmic-ray nucleon & $-$                  & 
$-$                  & $\pm$15.0          & $-$                & $\pm$2.2 \\
  Prompt model uncertainty            & $-$                  & 
$-$                  & $-$                & $^{+31.6}_{-40.4}$ & 
$^{+12.6}_{-16.1}$ \rule[-2mm]{0mm}{6mm} \\ \hline
  Total                               & $\pm$0.4(stat.)      & 
$\pm$9.1(stat.)      & $\pm$9.8(stat.)    & $\pm$1.1(stat.)    & 
$\pm$4.5(stat.) \rule[-2mm]{0mm}{6mm} \\ 
                                      & $^{+13.6}_{-12.4}$(syst.) & 
$^{+54.5}_{-100}$(syst.) & $^{+80.5}_{-58.7}$(syst.) & 
$^{+55.0}_{-59.8}$(syst.) &$^{+49.3}_{-68.7}$(syst.) 
\rule[-2mm]{0mm}{4mm} \\ \hline \hline
\end{tabularx}

\end{center}
\end{table*} 

Table~\ref{tb:Sys} summarizes the statistical and systematic errors for 
signal, atmospheric muon and neutrino,
prompt atmospheric neutrino, and the total background.

One of the dominant sources of systematic uncertainties in the signal event 
rates 
is the error associated with the Cherenkov photon measurement, namely
the relationship between measured NPE and the energy of the charged particles. 
This is due to limitations in the understanding of detector sensitivities, 
photon propagation in the ice and the detector response to bright events 
which, for example, involves saturation effects of the DOMs. 
This uncertainty is estimated by calibrating the absolute sensitivity of the 
DOMs in the laboratory
and by measuring it {\it in-situ} using a light source co-deployed with the 
DOMs in the ice~\cite{icecubeEHE2010, icecubeEHE2011}.
The other uncertainties in the signal rates involve the relevant interactions 
of neutrinos and leptons produced during the propagation through the Earth.
For example, the Landau-Pomeranchuk-Migdal (LPM) 
effect~\cite{LPM, 1956PhRv..103.1811M} 
can be important since it elongates electromagnetic showers.
The elongated shower length is about 20--40 m for 1--10 EeV 
electrons~\cite{Gerhardt:2010bj},
thus still being comparable to the IceCube DOM separation of 17 m, and hence 
negligible.
Uncertainties due to other propagation effects are estimated as 
described in ~\cite{icecubeEHE2010}.

The uncertainty of the systematic shifts
of reconstructed zenith angles for cascade events causes a systematic error 
in the estimation of the signal neutrino passing rate. 
The effect is NPE dependent and thus energy dependent. 
We artificially vary the systematic zenith angle shift by different factors
to evaluate the resulting uncertainties. The complete randomization of 
zenith angles was found to bring the largest
reduction of the cascade event selection efficiency.
The reduction is 20.0\% for events with energies below 10 PeV, 8.5\% between 
10 and 100 PeV and 2.0\% above 100 PeV.
Since most of the cosmogenic neutrino signal (99.6\%) is expected above 10~PeV
and the present analysis is mostly sensitive to track events above 10~PeV as 
seen in Fig.~\ref{fig:effarea},
the effect on the cosmogenic neutrino signal rate is quite limited.
The systematic error on the overall signal
rate due to the limited performance of the cascade event reconstruction is 
estimated to be $-$0.5\%.

Systematic errors in the atmospheric muon background rate arise from 
uncertainties in the primary cosmic-ray composition, 
the hadronic interaction model implemented in the air shower simulation, and 
the cosmic-ray flux variation at the relevant energies.
The two extreme cases of the cosmic-ray compositions, pure iron and pure 
proton, are used.
In the current analysis, the iron-only hypothesis is used for the baseline 
background rates. This leads to a higher, i.e. conservative, estimate of the 
photon yield from the muon bundles induced by primary cosmic-ray particles 
at a particular energy.
The difference between the pure-iron and the pure-proton hypothesis then 
provides the size of the relevant systematic uncertainty.
The uncertainty associated with the hadronic interaction model is estimated 
by switching the model from SIBYLL 2.1~\cite{Ahn:2009wx}
to QGSJET-II-03~\cite{Ostapchenko:2010vb} in the simulations. 
The uncertainty in the cosmic-ray flux normalization is estimated from the 
variance in the flux measured by several 
experiments~\cite{augerSpectrum2010,hiresSpectrum2008} relative to the one 
used in this analysis~\cite{Nagano:2000ve} at 10~EeV, the peak energy of 
primary cosmic rays that produce atmospheric muon events passing the final 
selection criteria. The contribution of the cosmic-ray normalization to the 
uncertainty in the atmospheric neutrino rate is estimated in a similar way 
at energies from 1 to 100\,PeV from various models~\cite{TG,Zatsepin2006}. 
In addition, a systematic uncertainty for the atmospheric neutrino rate 
arises from the uncertainty of 
the parametrization of the neutrino multiplicity as described in 
section~\ref{sec:datasimulation}.
A comparison to the full simulation by CORSIKA~\cite{Heck:1998vt} provides 
the relevant uncertainty. 
The systematic uncertainties for backgrounds associated with
the photon detection efficiency and the optical properties of the ice are 
determined in the same manner as for signal events.

The atmospheric neutrino background is calculated over $4\pi$ solid angle 
and simulated independently of the atmospheric muon background. In reality, 
downward-going atmospheric neutrino events would be accompanied by 
atmospheric muons, which improves their geometrical reconstruction.
Because correctly reconstructed downward-going events are mostly rejected 
due to the higher NPE threshold employed in the final event selection, 
the background rate obtained from the independent neutrino and muon 
simulations is likely overestimated.

The systematic error on the prompt neutrino flux is estimated similarly. 
Relatively large uncertainty arise from the parametrization in the framework 
of the Enberg {\it el al.} model~\cite{prompt_enberg} which we used for the 
calculation of the baseline rate of prompt neutrinos.
A possible non-perturbative QCD contribution in charm production involves an 
even larger uncertainty. We have not observed clear evidence for prompt 
contributions in atmospheric neutrinos so far \cite{Schukraft:2013ya}.

%

\section{\label{sec:Resuts} Results}

\begin{figure}[bt]
  \includegraphics[width=0.45\textwidth]{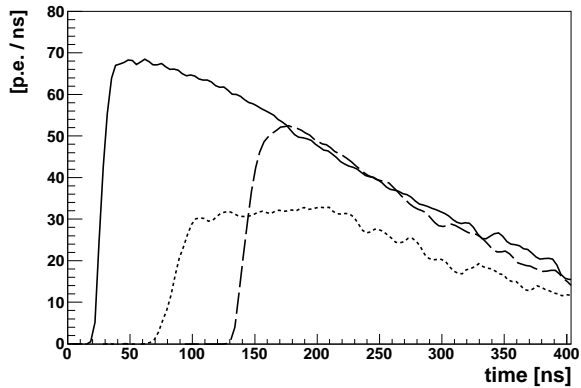}
  \caption{Waveforms of PMT outputs captured by three DOMs in the neighborhood 
    of the reconstructed vertex position of the event obtained in January, 
    2012. 
    The waveform drawn as a solid curve is recorded in the DOM closest to the 
    vertex (the brightest DOM).
    The waveform in the lower (upper) next-to-nearest to the brightest DOM on 
    the same string is shown as a dashed (dotted) curve.
    Photons arrive earlier in the upper DOM because it is closer to the 
    cascade vertex than the lower DOM.
    The signals from the upper DOM exhibit clear signatures of scattered 
    late photons, suggesting that this cascade is a downward-going event.
  }
\label{fig:waveform}
\end{figure}

Two events passing the final selection criteria are 
observed~\cite{Aartsen:2013bka}.
The waveform profiles and the detector hit patterns of both events are 
consistent with that of Cherenkov photons from particle cascades induced by 
neutrinos well inside the IceCube instrumentation volume. There is no 
indication of outgoing/incoming muon or tau tracks. Several waveforms 
captured by the DOMs in the neighborhood of one of the reconstructed cascade 
vertex position are shown in Fig.~\ref{fig:waveform}. The total charge 
contained in the waveforms plays a dominant role in estimating the deposited 
energy of the cascade. The leading edge time mainly determines the vertex 
position. 
The relative widths of the waveforms in DOMs in the forward an rear direction 
of the cascade is relevant for the reconstruction of the arrival direction 
of neutrinos. Since photons can only reach the backward direction by 
scattering, the distribution of photon arrival times is much wider in the 
backward region of the cascades. The relations of the waveform features to 
the energy, direction, and vertex position are described using a single 
likelihood function built from a product of Poisson probabilities of the 
number of photons predicted to arrive in a given time bin against the 
number extracted from the recorded waveform. Minimizing the log likelihood 
under simultaneous variation of the energy and geometry of the cascade 
hypothesis yields estimates of the deposited energy, direction, and 
interaction vertex of the cascade.

The reconstructed deposited energies of the two observed cascades 
are $1.04\pm 0.16$\,PeV and $1.14\pm 0.17$\,PeV, respectively. 
The statistical energy resolutions for these events are obtained by 
simulating cascades with parameters close to the reconstructed energies 
and cascade vertices,
and are found to be 3\%. The total error on the energy is dominated by 
systematic uncertainties. These  include the absolute detection efficiency 
of the DOM and the optical properties of the ice, both of which are major 
factors when relating the number of observed photons to the cascade energy.
The size of the errors are estimated by reconstructing simulated events with 
various models of the ice properties.

\begin{table}[bt]
\caption{The 90\% C.L. of the energy range of the primary neutrino in 
  PeV at the Earth's surface for the two events for an energy spectrum 
  following an E$^{-2}$ power law.}
\label{table:energyRange}
\begin{center}
\begin{tabular}{cc}
\hline
\hline
  &  Energy range ($90\%$ C.L.)\\
\hline
\hline
Event (August, 2011) & $0.81-7.6$ PeV\\
Event (January, 2012) & $0.93-8.9$ PeV\\
\hline
\hline
\end{tabular}
\end{center}
\end{table}

The incoming neutrino energy corresponds exactly to the deposited cascade 
energy if a charged-current interaction of an electron neutrino induces a 
cascade. 
For neutral current reactions of neutrinos of any flavor, only a fraction 
of the neutrino energy is transferred to a cascade depending on the 
inelasticity of the collision. Because the present analysis is 
incapable of distinguishing between neutrino flavors, both interaction 
channels are included when constructing the probability density function (PDF)
of the energy of the incoming neutrino. Here, the systematic uncertainties 
for the deposited energies are taken into account. 
The PDF of the neutrino energy at the surface of the Earth is built by 
simulating neutrino interactions over a wide energy range each time evaluating
the probability that the resulting cascade energy matches the estimated 
energy and its uncertainty.
The 90\% C.L.\ energy ranges obtained from the PDFs for neutrino spectra 
with an $E_\nu^{-2}$ power law flux
are summarized in Table~\ref{table:energyRange}. 
The flavor ratio is assumed to be $\nu_e:\nu_\mu:\nu_\tau= 1:1:1$.
Since the neutrino-nucleon interaction cross section increases with neutrino 
energy, the possibility that 
the energy of the primary neutrino is much higher than the observed cascade 
energy is not entirely negligible, depending on the neutrino spectrum.
For example, the 90\% C.L. energy range for a cosmogenic neutrino 
model~\cite{ahlers2010}
extends to about 500 PeV, which shows that the energy range heavily 
depends on the shape of the energy spectrum.

\section{\label{sec:ModelTests} Tests on cosmogenic neutrino models}

Our results are characterized by two observational facts:
the detection of two neutrinos with deposited energies of about one PeV
and the non-detection of neutrinos with higher deposited energies. 
First, we investigate whether a single cosmogenic neutrino model can account 
for these two observational facts simultaneously. Secondly, we constrain the 
UHECR origin with the present results. 
Because most cosmogenic neutrinos have energies above 100\,PeV, 
tests on the event rate above this energy expected from cosmogenic neutrino 
models under various assumptions on the UHECR spectrum and the evolutions of 
the source distributions will lead to constrains on the UHECR origin. 
We note that the energy threshold of 100\,PeV is an a posteriori parameter 
and, hence, the results are not part of the blind analysis.

The statistical significance of these tests is limited by our observational 
exposure. To obtain the best constraints, we combine the exposure of the 
previously published results obtained by the half-completed IceCube detector 
with its 40 string configuration (IC40)~\cite{icecubeEHE2011} with the present
results hereafter. The IC40 data increases the observational exposure by about
30\%, depending on the neutrino energy, as displayed in Fig.~\ref{fig:effarea}.

\subsection{\label{subsec:ModelTestsAllEnergy} The full energy range test}
We introduce here an energy inclusive test
which checks the consistency of the energy distributions of cosmogenic 
neutrino models with the observed two events.
A p-value is calculated with the Kolmogorov-Smirnov test (KS-test) using the 
energy spectrum of the neutrino models and the energy PDFs of the two observed
events. 
The expected energy distributions from the neutrino models are obtained by 
multiplication of the neutrino effective area with the predicted neutrino 
energy spectrum. This allows us to analytically calculate p-values without 
relying on extensive Monte-Carlo simulations.
In order to evaluate the final p-value, $P_E$, that the two events 
(energies $E_1$ and $E_2$) are consistent with a cosmogenic flux model, 
the p-value obtained in the KS-test, $P_{\rm KS}(E_1,E_2)$, is convoluted 
with the energy PDFs of the two events as follows:
\begin{equation}
P_{\rm E} = \int dE_1\rho_1(E_1)\int dE_2\rho_2(E_2) P_{\rm KS}(E_1,E_2),
\label{eq:KS}
\end{equation}
where $\rho_i$ is the energy PDF of the $i$th event.
Note that the PDF is different for each model to be tested as described in 
the previous section.
Table~\ref{table:CLwithTheTwoEvents} summarizes the resulting p-values of this
test: all cosmogenic neutrino models are inconsistent with the two observed 
events at more than 90\% C.L.

The recent follow-up analysis~\cite{IPA2013} revealed the existence
of neutrinos at TeV energies above the atmospheric background,
in addition to the two PeV events reported in Ref.~\cite{Aartsen:2013bka}. 
The event distribution indicated either a substantially softer spectrum than 
$E_\nu^{-2}$ or the presence 
of a break or cut-off at PeV energies, although the statistics are limited.
The present analysis confirmed this picture using the KS test with
an $E_\nu^{-2}$ spectrum hypothesis
as Table~\ref{table:CLwithTheTwoEvents} lists the resultant p-values
with various assumptions of the spectral cut-off energies. 
The observed PeV events are unlikely to originate from a bulk of neutrinos 
with energies extending
well above PeV, regardless of the characteristics of the events at TeV 
energies found in the follow-up analysis.

\begin{table}[bt]
\caption{P-values $P_{\rm E}$ in Eq.~\ref{eq:KS} are listed for several 
neutrino models.
All the models shown here assume the cosmic-ray primaries to be protons and
different spectral indices/cut-off energies at sources, IR/UV backgrounds, 
as well as different cosmological evolution parameters and extension in
redshift for the sources. 
P-values for $E_{\nu}^{-2}$ spectra with various cut-off energies
are also shown for reference.}
\label{table:CLwithTheTwoEvents}
\begin{center}
\begin{tabular}{lc}
\hline
\hline
$\nu$ Model & p-value \\\hline
Yoshida and Teshima~\cite{yoshida93}     & \\
$m = 4.0,z_{\rm max}=4.0$                    & 0.077 \\
\hline
Ahlers {\it et al.}~\cite{ahlers2010}    & \\
$m = 4.6,z_{\rm max}=2.0$ (``the best fit'') & 0.075\\
\hline
Kotera {\it et al.}~\cite{kotera2010}    & \\
GRB                                      & 0.052 \\ 
\hline
Kotera {\it et al.}~\cite{kotera2010}    & \\
Fanaroff-Riley type II                   & 0.039 \\ 
\hline
$E_{\nu}^{-2}$ &      \\
with cut-off at 10  PeV                  & 0.18  \\ 
with cut-off at 100 PeV                  & 0.13  \\ 
with cut-off at 1 EeV                 & 0.11  \\ 
\hline
\hline
\end{tabular}
\end{center}
\end{table}

\subsection{\label{subsec:ModelTestsAbove100PeV} The ex post facto test 
above 100\,PeV}

Here, a prospective event rate in the energy region above 100\,PeV is compared
to the observational upper limit. A constraint on a given neutrino model is 
set by calculating the model rejection factor (MRF)~\cite{MRF} given by
\begin{equation}
R_{\rm MRF} = \frac{N_{100(1-\alpha)\%}}{\mu_{\nu}},
\label{eq:MRF}
\end{equation}
where $N_{100(1-\alpha)\%}$ is the upper limit of number of events at 
$100(1-\alpha) \%$ C.L. and $\mu_{\nu}$ is the event rate of signal 
neutrino events predicted by the model above 100\,PeV. Any model with 
$R_{\rm MRF} \leq 1$  is rejected at $\geq 100(1-\alpha)\%$ C.L.\ in this 
approach. 
For $\alpha = 0.1$, $N_{90\%} = 2.27$ in the Feldman-Cousins 
approach~\cite{feldman98}
for a null observation with a conventional background of 0.16 events.
The large number of background events comes mainly from the IC40 analysis 
contributing 0.11 event~\cite{icecubeEHE2011}.
Although the probability that the original neutrino energy of the two observed
 events is higher than 100\,PeV is expected to be small, this is taken into 
account by calculating the most probable upper limit:
\begin{equation}
N_{100(1-\alpha)\%} = \sum\limits_{n=0}^{2}P_n N_{100(1-\alpha)\%}^n.
\label{eq:UL}
\end{equation}
Here, $P_n$ is the probability of finding $n$ events above 100\,PeV determined 
by 
the energy PDFs of the two events, and $N_{100(1-\alpha)\%}^{n}$ is the upper 
limit for $n$ observed events. 
Since the energy PDF highly depends on the shape of the energy spectrum, 
an appropriate shape of an energy spectrum has to be chosen.
Since the two observed events were found to be inconsistent with cosmogenic 
neutrino models 
as shown in the previous sub-section, the cosmogenic neutrino models are not 
used for the energy PDF, instead an $E^{-2}$ power law spectrum is used.
The $N_{90\%}$ is calculated for the standard cosmogenic models,
and found to be 2.273 which is slightly larger than for the case of a null 
detection.
The systematic uncertainty on the background estimates is incorporated using 
a method outlined in~\cite{tegenfeldt2005}. The p-value $\alpha$ for a given 
model is obtained by requesting $R_{\rm MRF}=1$ in Eq.~\ref{eq:MRF}.

\begin{table}[bt]
\caption{Expected numbers of events from several neutrino models
and the p-values for consistency with the present observation in energy range
above 100 PeV.}
\label{table:CL}
\begin{center}
\begin{tabular}{lcc}
\hline
\hline
$\nu$ Model & Event rate above 100 PeV & p-value \\
\hline
Yoshida and Teshima~\cite{yoshida93} &    & \\
$m = 4.0,z_{\rm max}=4.0$&  2.0 &  0.14 \\
\hline 
Kalashev {\it et al.}~\cite{kalashev02} &   &  \\
$m = 5.0,z_{\rm max}=3.0$& 3.1 &  0.045 \\
\hline
Yoshida and Ishihara~\cite{yoshida2012} &    & \\
$m = 5.0,z_{\rm max}=2.0$&  1.5 &  0.22 \\
\hline
Ahlers {\it et al.}~\cite{ahlers2010} &    & \\
$m = 4.6,z_{\rm max}=2.0$&  1.5  & 0.22 \\
("the best fit") &  & \\
\hline
Ahlers {\it et al.}~\cite{ahlers2010} &    & \\
("the maximal flux") & 3.1  & 0.044 \\
\hline
Kotera {\it et al.}~\cite{kotera2010} &    & \\
GRB & 0.48 & 0.66 \\ 
\hline
Kotera {\it et al.}~\cite{kotera2010} &    & \\
SFR & 0.46 & 0.67 \\ 
\hline
Kotera {\it et al.}~\cite{kotera2010} &    & \\
Fanaroff-Riley type II  & 2.9 &  0.052 \\ 
\hline
Top-down 1~\cite{sigl99} &  & \\
SUSY    & 16 & $\leq 0.0020$ \\
\hline
Top-down 2~\cite{sigl99} &  & \\
GUT    & 3.9 & 0.021 \\
\hline
\hline
\end{tabular}
\end{center}
\end{table}

\begin{figure}[tb]
  \includegraphics[width=0.4\textwidth]{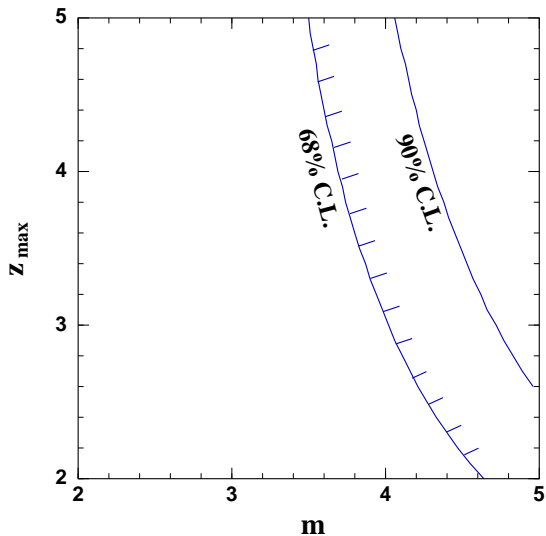}
  \caption{Constraints on the UHECR source evolution parameters of 
    $m$ and $z_{\rm max}$ 
    with the present analysis. The semi-analytic formulation~\cite{yoshida2012}
    estimates the neutrino flux for calculating the limit shown here.
    The area above the solid lines is excluded at the quoted confidence level.
 \label{fig:gzk_constraint}}
\end{figure}

\begin{figure}[bt]
  \includegraphics[width=0.45\textwidth]{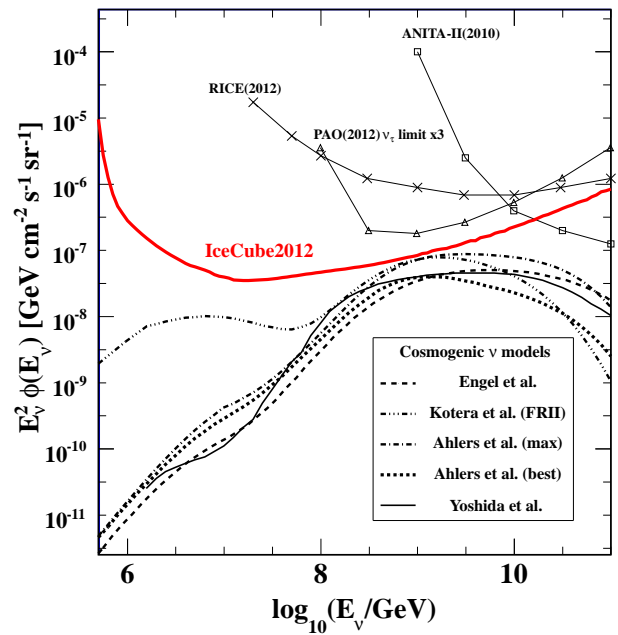}
  \caption{All flavor neutrino flux differential 90\% C.L. upper limit 
    evaluated for each energy with a sliding window of one energy decade
    from the present IceCube EHE analysis including the IceCube exposure from 
    the previously published result (IC40)~\cite{icecubeEHE2011}.
    All the systematic errors are included. Various model predictions 
    (assuming primary protons)
    are shown for comparison; Engel {\it et al.}~\cite{ESS},
    Kotera {\it et al.}~\cite{kotera2010}, 
    Ahlers {\it et al.}~\cite{ahlers2010}, 
    Yoshida {\it et al.}~\cite{yoshida93}.
    The model-independent differential 90\% C.L.
    upper limits for one energy decade by other experiments are also shown for 
    Auger (PAO)~\cite{auger12}, RICE~\cite{rice11}, 
    ANITA~\cite{anita2,anita2err} with appropriate
    normalization by taking into account the energy bin width and the neutrino 
    flavor. The upper limit for the $\nu_\tau$ flux obtained by Auger is
    multiplied by three to convert it to an all flavor neutrino flux limit 
    (assuming an equal neutrino flavor ratio).
 \label{fig:UL}} 
\end{figure}

Table~\ref{table:CL} summarizes the p-values for several neutrino models.
The maximal flux allowed by the constraints from the diffuse photon flux 
(labeled as ``the maximal flux'' in the table) is excluded at 95\% C.L. 
It demonstrates that the present constraints from the limit on the ultra-high 
energy neutrino flux are compatible with those from photon flux measurements 
by Fermi in the 10 GeV region~\cite{fermilimit}.

In order to set constraints on characteristics of the UHECR sources in a more 
comprehensive manner,
a parametrization often used in the literature~\cite{yoshida93} is employed, 
in which the spectral emission rate per co-moving volume scales as 
$(1+z)^m$ for $z\leq z_{\rm max}$.
The event rate at energies above 100 PeV is calculated for a given $m$, 
and $z_{\rm max}$
using the formula in Ref.~\cite{yoshida2012}.
The constraints on the parameter space of $m$ and $z_{\rm max}$ are derived by 
using Eq.~\ref{eq:MRF}, and are displayed in Fig.~\ref{fig:gzk_constraint}.

\subsection{\label{subsec:DiscussionsOnTheTests} Discussion}

The models listed in the top two rows of Table~\ref{table:CLwithTheTwoEvents} 
assume that the {\it ankle} structure which appears at 3 to 10\,EeV in the 
UHECR spectrum
is due to the transition from the Galactic to the extragalactic 
component~\cite{wibig2005}. 
In this scenario, the cosmogenic neutrino generation mechanism is 
dominated by collisions of UHECRs with the CMB photons which results in a 
neutrino energy spectrum with a peak at about 1~EeV, well above the main 
regime of the energy range of the two observed events.
This is the reason why these models are inconsistent with the two observed 
events as shown in Table~\ref{table:CLwithTheTwoEvents}.
The models in the lower two rows of Table~\ref{table:CLwithTheTwoEvents} 
(Kotera {\it et al.}~\cite{kotera2010}) assume the ``dip'' transition 
model~\cite{berezinsky2006} where the ankle structure is mainly caused by 
pair-production energy losses of UHECRs on diffuse infrared, optical, and 
ultraviolet backgrounds (IR/UV backgrounds) during intergalactic propagation.
The neutrino models in Kotera {\it et al.}~use the IR/UV backgrounds as 
modeled by Stecker~\cite{stecker2006} which comprises an increased 
far-infrared bump at large redshift 
(note that the IR/UV model employed in these neutrino models is now 
disfavored by gamma-ray observation with Fermi-LAT~\cite{Ackermann:2012sza}). 
Compared to the standard cosmogenic models, the dip and the IR/UV 
backgrounds leads to an increased flux of neutrinos at PeV energies, 
so that these models in Kotera {\it et al.} could be more consistent with 
the observation.
However, even in these models, the collision of UHECRs with CMB photons 
produces a bulk of neutrinos with energies much higher than 100 PeV which 
should have been detected 
because of the significantly larger effective area at these energies. 
In addition, the substantial flux at PeV energies yields energy PDFs for the 
observed two events very similar to those from an $E_\nu^{-2}$ spectrum.
Since the energy range for the $E_\nu^{-2}$ spectrum PDF does not extend to 
10 PeV as shown in Table~\ref{table:energyRange},
neutrinos with energy of 100 PeV or greater are less likely to be 
responsible for the observed PeV cascades.
Because of these reasons,  p-values for these scenarios in 
Kotera {\it et al.} are small as shown in Table~\ref{table:CLwithTheTwoEvents}.
In conclusion, none of the cosmogenic scenarios is consistent with the 
observation of the two events.
This indicates that models which predict neutrino spectra extending to 
energies well beyond 100 PeV will not explain our measurements. 

The model test based on the event rates above 100\,PeV indicates that strong 
source evolution models ($m\gg 4$) are not responsible for the bulk of UHECRs.
Among sources categorized in this class are the Fanaroff-Riley type II (FR-II)
radio galaxies, the long-standing favorite as a candidate of the UHECR 
emitters~\cite{biermann87}. 
Similarly a strong source evolution model for GRBs~\cite{Yuksel:2006qb} is 
also rejected by our observation
since the model produces higher neutrino flux than the FR-II model.
The obtained limits are highly complementary to the bound from the diffuse 
photon flux~\cite{fermilimit}, because the cosmogenic neutrino intensity 
around 1\,EeV, the central energy range of the presented search with IceCube, 
is stable against uncertainties in the IR/UV backgrounds and the transition 
model between the galactic and extragalactic component of the 
UHECRs~\cite{kotera2010,yoshida2012,takami09,decerprit2011}. 
We should note, however, that the obtained bound is not valid if the mass 
composition of UHECRs is not dominated by proton primaries. 
The dominance of proton primaries is widely assumed in the models mentioned 
here while a dominance of heavier nuclei such as iron provides at least 2--3 
times lower neutrino fluxes. The analysis is not sensitive enough to 
reach these fluxes yet.

\section{\label{sec:ModelIndepUpLimit} The Model independent upper limit}
The quasi-differential, model-independent 90\% C.L. upper limit on all flavor 
neutrino fluxes, $\phi_{\nu_e + \nu_\mu + \nu_\tau}$, 
was evaluated for each energy with a sliding window of one energy decade.
It is shown in Fig.~\ref{fig:UL} 
using the same method as implemented in our previous EHE neutrino 
searches~\cite{icecubeEHE2010,icecubeEHE2011}. 
An equal flavor ratio of $\nu_e:\nu_\mu:\nu_\tau = 1:1:1$ is assumed here.
A difference from the calculation of the limit shown in our previous 
publications arises from the existence of two events in the final sample. 
The 90\% event upper limit used in the calculation takes into account the 
energy PDFs of each of the observed events using  Eq.~\ref{eq:UL}, 
where $P_n$ is a function of the neutrino energy $E_\nu$ and corresponds to 
the probability of having $n$ events in the interval 
$[\log_{10}(E_\nu/{\rm GeV})-0.5,\ \log_{10}(E_\nu/{\rm GeV})+0.5]$. 
Here, the PDFs for an $E_{\nu}^{-2}$ spectrum are used since the two observed 
events are not consistent with a harder spectrum such as from cosmogenic 
neutrino models.
The quasi-differential limit takes into account all the systematic 
uncertainties described in section~\ref{sec:Sys}.
The effect of the uncertainty due to the angular shift of the cascade 
events on the upper limit is negligible above 10~PeV 
($< 1\%$) as track events dominate in this energy range.
Below 10 PeV, the effect weakens the upper limit by 17\% because cascade 
events dominate.
Other systematic uncertainties are implemented as in previous EHE neutrino 
searches~\cite{icecubeEHE2010,icecubeEHE2011}.
The obtained upper limit is the strongest constraint in the EeV regime so far.
In the PeV region, the constraint is weaker due to the detection of the 
two events. 
An upper limit for an $E^{-2}$ spectrum that takes into account the two 
observed events
was also derived and amounts to $E^2 \phi_{\nu_e + \nu_\mu + \nu_\tau}
= 2.5 \times 10^{-8}$ GeV cm$^{-2}$ s$^{-1}$ sr$^{-1}$
for an energy range of 1.6~PeV -- 3.5~EeV (90\% event coverage).

\section{\label{sec:summary} Summary}
We analyzed the 2010-12 data samples collected by the 79 and 86-string 
IceCube detector searching for extremely high energy neutrinos with energies 
exceeding 1\,PeV. We observed two neutrino-induced cascade events passing the 
final selection criteria. The energy profiles of the two events indicate 
that these events are cascades with deposited energies of about 1\,PeV.
The cosmogenic neutrino production is unlikely to be responsible for these 
events. An upper limit on the neutrino rate in the energy region above 
100\,PeV places constraints on the redshift distribution of UHECR sources. 
For the first time the observational constraints reach the flux region 
predicted for some UHECR source class candidates. 
The obtained upper limit is significantly stronger compared to our previous 
publication~\cite{icecubeEHE2011} because of the enlarged instrumented volume 
and the refined Monte Carlo simulations. Future data obtained with the 
completed detector will further enhance IceCube's sensitivity to 
cosmogenic neutrino models.

\begin{acknowledgments}

We acknowledge the support from the following agencies:
U.S. National Science Foundation-Office of Polar Programs,
U.S. National Science Foundation-Physics Division,
University of Wisconsin Alumni Research Foundation,
the Grid Laboratory Of Wisconsin (GLOW) grid infrastructure at the 
University of Wisconsin - Madison, the Open Science Grid (OSG) 
grid infrastructure;
U.S. Department of Energy, and National Energy Research Scientific 
Computing Center,
the Louisiana Optical Network Initiative (LONI) grid computing resources;
Natural Sciences and Engineering Research Council of Canada,
WestGrid and Compute/Calcul Canada;
Swedish Research Council,
Swedish Polar Research Secretariat,
Swedish National Infrastructure for Computing (SNIC),
and Knut and Alice Wallenberg Foundation, Sweden;
German Ministry for Education and Research (BMBF),
Deutsche Forschungsgemeinschaft (DFG),
Helmholtz Alliance for Astroparticle Physics (HAP),
Research Department of Plasmas with Complex Interactions (Bochum), Germany;
Fund for Scientific Research (FNRS-FWO),
FWO Odysseus programme,
Flanders Institute to encourage scientific and technological research in 
industry (IWT),
Belgian Federal Science Policy Office (Belspo);
University of Oxford, United Kingdom;
Marsden Fund, New Zealand;
Australian Research Council;
Japan Society for Promotion of Science (JSPS);
the Swiss National Science Foundation (SNSF), Switzerland;
National Research Foundation of Korea (NRF)

\end{acknowledgments}


\end{document}